\documentclass[twocolumn,reprint,aps]{revtex4-1}
\usepackage{amsmath,amssymb,amsthm,bm,braket,ascmac,bbm}
\usepackage[dvipdfmx]{graphicx}
\usepackage{color}

\usepackage{docmute}

\usepackage{comment}
\newcommand{\I}{\hat{I}}
\newcommand{\J}{\hat{J}}
\newcommand{\Id}{\hat{I}^{d}}

\newcommand{\hatrho}{\hat{\rho}}
\newcommand{\tr}{\text{Tr}}

\newcommand{\hsinner}[2]{\braket{#1,#2}}

\newcommand{\hH}{\hat{H}}
\newcommand{\hP}{\hat{P}}
\newcommand{\hQ}{\hat{Q}}
\newcommand{\hO}{\hat{A}}
\newcommand{\hbO}{\hat{\overline{A}}}
\newcommand{\dO}{\delta\hat{A}}
\newcommand{\hC}{\hat{C}}

\newcommand{\Lin}{L_{\mathrm{ini}}}
\newcommand{\ave}[2]{\braket{#1}_\text{#2}}

\newcommand{\hl}[1]{\textcolor{black}{#1}}
\newcommand{\hlb}[1]{\textcolor{black}{#1}}
\newcommand{\hlm}[1]{\textcolor{black}{#1}}

\newcommand{\hlf}[1]{\textcolor{black}{#1}}
\newcommand{\hlik}[1]{\textcolor{black}{#1}}

\begin{document}

%\title{An unthe DErlying mechanism of validity of generalized Gibbs ensembles }
%\title{Justification of generalized Gibbs ensemble from eigenstate-thermalization-hypothesis viewpoint}
\title{Noncommutative generalized Gibbs ensemble in isolated integrable quantum systems}

\author{Kouhei Fukai}
\email{k.fukai@issp.u-tokyo.ac.jp}
\affiliation{The Institute for Solid State Physics, The University of Tokyo, Kashiwa, Chiba 277-8581, Japan}

\author{Yuji Nozawa}
\affiliation{The Institute for Solid State Physics, The University of Tokyo, Kashiwa, Chiba 277-8581, Japan}

\author{Koji Kawahara}
\affiliation{The Institute for Solid State Physics, The University of Tokyo, Kashiwa, Chiba 277-8581, Japan}

\author{Tatsuhiko N. Ikeda}
\email{tikeda@issp.u-tokyo.ac.jp}
\affiliation{The Institute for Solid State Physics, The University of Tokyo, Kashiwa, Chiba 277-8581, Japan}

%#### data & abstract ####
\date{\today}
\begin{abstract}
The generalized Gibbs ensemble (GGE), which involves multiple conserved quantities other than the Hamiltonian, has served as the statistical-mechanical description of the long-time behavior for several isolated integrable quantum systems. The GGE may involve a noncommutative set of conserved quantities in view of the maximum entropy principle, and show that the GGE thus generalized (noncommutative GGE, NCGGE) gives a more qualitatively accurate description of the long-time behaviors than that of the conventional GGE. Providing a clear understanding of why the (NC)GGE well describes the long-time behaviors, we construct, for noninteracting models, the exact NCGGE that describes the long-time behaviors without an error even at finite system size. It is noteworthy that the NCGGE involves nonlocal conserved quantities, which can be necessary for describing long-time behaviors of local observables.
\hlf{We also give some extensions of the NCGGE and demonstrate how accurately they describe the long-time behaviors of few-body observables.} 
\end{abstract}
\maketitle

%##########################################################
%######## \UTF{2460}イントロ  ##################################
%##########################################################

\section{Introduction}
%大目標と冷却原子との関連
%孤立量子系の緩和が調べられ、多くの場合で熱平行状態に落ち着くことがわかってきた。
The foundation of quantum statistical mechanics has seen a resurgence of interest
in recent years~\cite{DAlessio2016,Eisert2015a,Ilievski_2016,Mori_2018} partly because well-isolated and -controlled
artificial quantum systems have emerged as the ideal platform
to reconsider the long-standing problem~\cite{Kinoshita2006,Trotzky2011,Langen2013,Langen2015,Schreiber2015,Kaufman2016}.
A remarkable finding is that an isolated quantum many-body system
can relax to an effective stationary state even without energy dissipation or quantum decoherence.
\hlb{Although the stationary state in the strict sense appears only in infinite systems,
an effective (or approximate) stationary state arises
at large but finite system sizes, where the fluctuations and recurrences are negligible~\cite{Tasaki1998,Reimann2008,Linden2009a}.}

In generic nonintegrable systems, the effective stationary state coincides in fact
with the thermal state due to the eigenstate thermalization hypothesis (ETH)~\cite{Deutsch1991,Srednicki1994a},
which dates back to von Neumann~\cite{Neumann2010} and has recently been numerically verified~\cite{Rigol2008,Santos2010,Kim2014,Beugeling2014,Steinigeweg2014b,Khodja2015,Yoshizawa2018}.
Meanwhile, there exist known systems in which the stationary state does not coincide with
the thermal state such as integrable systems~\cite{Rigol2007,Santos2010,Biroli2010,Ikeda2013,Alba2015}, many-body localized systems~\cite{Pal2010,Iyer2013}, and so on~\cite{Worm2013,Shiraishi2017,Mondaini2018}.
It remains an open question how to classify all the nonthermal systems
and to identify the statistical-mechanical ensemble describing those states.

%#####
%conserved quantitiesのダイナミクス への影響、可積分系と言う特異例の存在
The generalized Gibbs ensemble (GGE) is a paradigmatic framework
to describe various nonthermal stationary states~\cite{Rigol2007}.
Whereas the usual Gibbs (canonical) ensemble involves the Hamiltonian,
the GGE does other conserved quantities as well (see Eq.~\eqref{eq:gge} below)~\cite{Jaynes1957}.
The GGE describes the stationary states in noninteracting integrable models
(hard-core bosons~\cite{Rigol2007}, the transverse-field Ising model~\cite{Vidmar_2016}),
interacting (Bethe-ansatz) integrable ones~\cite{Caux2012a,Pozsgay_2013, PhysRevB.89.125101, PhysRevLett.113.117202,PhysRevLett.113.117203,PhysRevLett.113.117202,PhysRevLett.115.157201},
models with different-type conserved quantities~\cite{Hamazaki2016},
quantum field theories~\cite{Cardy_2016}, and so on~\cite{PhysRevA.78.013626,Kollar2011}.

%##########################################################
%######## \UTF{2461}GGEが失敗する例もあることをいう  ##################################
%##########################################################
Despite its success, the GGE sometimes fails to describe the stationary state.
%localizationとGGE失敗の関係
For example, Spinless fermions or hard-core bosons under incommensurate potential cannot be described by the GGE
due to the localization of single particle eigenstate~\cite{Caneva_2011,PhysRevA.87.063637,PhysRevLett.109.247205,PhysRevB.87.064201,2017JSMTE..11.3107W}.
Another example is the entanglement prethermalization in an interacting integrable system~\cite{Kaminishi2015},
where nonlocal conserved quantities play significant roles.
One crucial problem is that the GGE is a general framework
and never tells us which conserved quantities should be incorporated.
When a GGE fails, it is hard to tell whether the ad hoc set of conserved quantities is not enough
or the framework breaks down.
In particular,
the GGEs mentioned above implicitly assume that the conserved quantities commute with each other (commutative GGE, CGGE), 
and this assumption may unnecessarily constrain the GGE. the GGE conserved quantities can be noncommutative
in view of the maximum entropy principle.
%~\hl{(see, however, Refs.~\cite{Fagotti_2014,YUKALOV20112797} for earlier discussions)}.
\hlb{The GGE with a noncommutative set of conserved quantities was first introduced in Ref.~\cite{Fagotti_2014}
in discussing the prerelaxation for the XY spin chain.
%from the superintegrable to integrable which does not have the additional local conservation lows is concentrated on in the XY chain.
The ensemble with a noncommutative set of conserved quantities was also mentioned in Ref.~\cite{YUKALOV20112797}.
However, it has not been systematically studied why and how those GGEs describe local or few-body observables well.}

%##########################################################
%######## \UTF{2462}以上を踏まえて新しいGGEを今回主張すると主張する  #########################
%##########################################################
\hlb{In this paper, we systematically study how the additional noncommutative conserved quantities affect the GGE}
and show that the GGE thus generalized (noncommutative GGE, NCGGE) describes the stationary states
in isolated integrable systems
better than the conventional CGGE.
By introducing the observable projection idea,
we provide a clear understanding of why the (NC)GGE
well describes the stationary states.
In this spirit,
for a noninteracting model,
we systematically construct the NCGGE
that describes the stationary states without an error at finite system size \hl{for \hlb{few}-body observables}.
\hlf{We also propose some extensions of the NCGGE and demonstrate how they work.}
%\hl{We have done the numerical calculation on the noninteracting integrable systems, but the concept of the observable projection idea and the uniqueness of the density matrix of the NCGGE are also valid in the interacting integrable systems.}

\hl{
The rest of this paper is organized as follows.
In Sec.~II, we formulate the problem and define the NCGGE. In Sec.~III, we explain why the GGE is valid with enough conserved quantities and the necessity of the NCGGE.
\hlb{The observable projection idea and the uniqueness of the NCGGE
presented in Secs.~II and III
are so general that they can be applied to both interacting and noninteracting integrable systems.
In Secs.~IV and V, focusing on free fermions, we show more detailed analyses of the NCGGE.}
}
\hl{In Sec.~IV, we give the example of NCGGE in free fermion and show the exactness of NCGGE even at finite system size. In Sec.~V, we give numerical results for two-body observables in the CGGE and NCGGE. In Sec.~VI, we give further extensions of the NCGGE.
Finally, in Sec.~VII, we summarize our study with concluding remarks.
}

%##################################
\section{Formulation of problem and NCGGE}
We consider an isolated quantum system described by a time-independent Hamiltonian $\hH$.
We let $\{E_n\}$ denote the distinct eigenenergies,
having $\hH=\sum_m E_m\hP_m$
with $\hP_m$ being the projection operator onto the corresponding eigenspace.
Under the Hamiltonian,
an initial state $\ket{\psi_{\text{ini}}}$ evolves as $\ket{\psi(t)}=e^{-iHt}\ket{\psi_{\text{ini}}}=\sum_m e^{-iE_mt}\hP_m\ket{\psi_\text{ini}}$ at time $t$
($\hbar=1$ throughout this paper).
Assuming that $\ket{\psi_{\text{ini}}}$ is a superposition of exponentially-large number (in terms of the system size) of energy eigenstates~\cite{Reimann2008,Short_2011,Reimann2012a},
we have an effective stationary state, in which
an observable $\hat{A}$ has its expectation value equal to the long-time average
\begin{align}\label{eq:lt}
\braket{\hat{A}}_{\mathrm{LT}}
=
\overline{\braket{\psi(t)|\hat{A}|\psi(t)}}
=\sum_m \braket{\psi_{\text{ini}}|\hP_m\hO\hP_m|\psi_{\text{ini}}},
\end{align}
where $\overline{f(t)}\equiv\lim_{T\rightarrow \infty}
\int_{0}^{T}(dt/T)f(t)$.
It is convenient to define the diagonal and off-diagonal decomposition of $\hO$ by
$\hO=\hbO+\dO$
with  $\hbO\equiv\sum_{m}\hP_m\hO\hP_m$ and $\dO\equiv\sum_{m,n\,(m\neq{n})}\hP_m\hO\hP_n$.
This notation simplifies Eq.~\eqref{eq:lt} as
\begin{align}\label{eq:diag}
	\braket{\hat{A}}_{\mathrm{LT}} = \braket{\hbO}_{\text{ini}}
	\equiv\braket{\psi_{\text{ini}}|\hbO|\psi_{\text{ini}}}.
\end{align}
If $\hat{A}$ is a conserved quantity $\hQ$, i.e. $[\hQ,\hH]=0$,
Eq.~\eqref{eq:lt} leads to
\begin{align}\label{eq:constraints}
\braket{\hQ}_{\mathrm{LT}}=\braket{\hat{Q}}_{\text{ini}}.
\end{align}

Equation~\eqref{eq:lt} gives $\braket{\hat{A}}_{\mathrm{LT}}$ exactly but involves an exponentially-large number
of inputs corresponding to every detail of $\ket{\psi_{\text{ini}}}$.
The question that we address in this paper is to find a statistical-mechanical ensemble $\rho$
which, \hlb{ with fewer (up to a polynomially-large number of) inputs},
satisfies $\braket{\hat{A}}_{\mathrm{LT}}\simeq\tr(\hatrho\hat{A})$ for \hl{local or few-body observables} $\hO$'s of interest.
\hlb{Here, $\simeq$ allows an error due to the finite-size effect
that vanishes in the thermodynamic limit. }

%##########################################################
%######## GGEのJeyesの原理からの導出  #########################
%##########################################################
The GGE is a successful candidate for such an ensemble formulated as follows.
The central idea is that the ensemble $\hatrho$ would
maximize the von Neumann entropy $S(\rho)=-\tr (\hatrho \log \hatrho)$ (the Boltzmann constant is set to unity).
When there exist multiple conserved quantities $\{\hat{Q}_{\alpha}\}$
including the Hamiltonian,
the dynamics is constrained by Eq.~\eqref{eq:constraints} for each $\hQ=\hQ_\alpha$.
Then the ensemble that maximizes the entropy under the constraints
is given by the stationary condition for 
\begin{align}\label{eq:psi}
\Psi(\hatrho, \{\lambda_{\alpha}\})=S(\hatrho)-\sum_{\alpha}\lambda_{\alpha}[\tr( \hatrho\hat{Q}_{\alpha})-\braket{\hat{Q}_{\alpha}}_{\text{ini}}],
\end{align}%
with the generalized temperatures $\{\lambda_\alpha\}$.
This condition leads to~\cite{Jaynes1957}
\begin{align}\label{eq:gge}
%\hat{\rho}_{\mathrm{GGE}}=Z^{-1} \exp \left[-\sum_{m} \lambda_{m} \hat{Q}_{m}\right]
\hat{\rho}_{\mathrm{GGE}}=\frac{e^{-\sum_{\alpha} \lambda_{\alpha} \hat{Q}_{\alpha}}}{Z},
\end{align}%
where 
$Z\equiv\tr\,e^{-\sum_{\alpha} \lambda_{\alpha} \hat{Q}_{\alpha}}$
is the partition function and the Lagrange multipliers $\{\lambda_{m}\}$
called the generalized temperatures are determined uniquely by
\begin{align}
\braket{\hat{Q}_{\alpha}}_{\text{ini}}=\tr(\hatrho_{\mathrm{GGE}}\hat{Q}_\alpha),
\end{align}
 for each $\alpha$. 
When $\{\hQ_\alpha\}$ consists only of the Hamiltonian,
the GGE reduces to the usual Gibbs (canonical) ensemble
and the generalized temperature is the inverse temperature $\beta$.
Once determined, the GGE gives expectation values for generic observables by
$\braket{\hO}_\text{GGE}\equiv\tr(\hatrho_{\mathrm{GGE}}\hat{A})$.

%#### NC #####
We emphasize that, in deriving Eq.~\eqref{eq:gge},
we never use the commutativity $[\hQ_\alpha,\hQ_\beta]=0$,
which is implicitly assumed in the literature.
In the Heisenberg model, for example,
the SU(2) symmetry implies that each of the total $S^x$, $S^y$,
and $S^z$ is a conserved quantity,
and one can construct the GGE by using all of them.
Thus, allowing noncommutative ones
increases the number of conserved quantities 
and improves the GGE in general.

%######
\hlb{We note that, when $[\hQ_\alpha,\hQ_\beta]\neq0$,
we cannot decompose Eq.~\eqref{eq:gge} into
the exponentials for each conserved quantity:
$e^{-\sum_{\alpha} \lambda_{\alpha} \hat{Q}_{\alpha}}\neq \prod_\alpha e^{-\lambda_\alpha \hat{Q}_\alpha}$.
Nevertheless, the exponential $e^{-\sum_{\alpha} \lambda_{\alpha} \hat{Q}_{\alpha}}$ is well-defined
and the generalized temperatures are uniquely determined.
We prove these facts in the Appendix~\ref{uniqueness}.}

%##########################################################
%######## \UTF{2463}ETHの論文との関連についてのセットアップ#########################
%##################################
\section{Validity of NCGGE in thermodynamic limit}
Before discussing concrete models,
we show why the GGE well describes the long-time behaviors~\eqref{eq:lt}
for generic observables in the thermodynamic limit.
Although the GGE is usually justified by the generalized ETH~\cite{Cassidy2011},
we here provide another perspective, in which the merit of the NCGGE becomes evident.

%####
To justify the GGE,
we invoke the observable projection with conserved quantities~\cite{PhysRevLett.124.040603}.
Note that the Hilbert-Schmidt inner product can be defined
between two \hlf{traceless} observables $\hat{A}$ and $\hat{B}$
as $\hsinner{\hat{A}}{\hat{B}}\equiv \tr (\hat{A}\hat{B})/D$ with $D$ being the Hilbert-space dimension. 
%\hlf{We \hlik{take an observable $\hat{A}$ of interest that is traceless and normalized as $0<\lim _{L \rightarrow \infty}\|\hat{A}\|<\infty$, where $L$ is the system size and $\|\hat{A}\|$ denotes the Hilbert-Schmidt norm,} $\|\hat{A}\|\equiv\sqrt{ \braket{\hat{A},\hat{A}}}$.}
For a given orthogonal set of conserved quantities $\{\hat{Q}_{\alpha}\}$,
we can decompose the observable $\hat{A}$ into the parallel and perpendicular components:
$\hat{A}=\hat{A}_{\parallel}+\hat{A}_{\perp}$,
where
$\hat{A}_{\parallel}=\sum_{\alpha} p_{A \alpha} \hat{Q}_{\alpha}$
and
$ p_{A \alpha}\equiv \hsinner{\hat{A}}{\hat{Q}_{\alpha}}/\hsinner{\hat{Q}_{\alpha}}{\hat{Q}_{\alpha}}$.
\hlf{We call  our $\{\hQ_\alpha\}$  a ``complete'' set of conserved quantities when the diagonal component $\hbO_{\perp}$, which is relevant in the long-time average, vanish in the thermodynamic limit.}
\begin{comment}
According to Ref.~\cite{PhysRevLett.124.040603},
\hlf{if $\lim_{L\rightarrow\infty}\|\hbO_{\perp}\|^{2}>0$ does not vanish in the thermo}
if our $\{\hQ_\alpha\}$ is a ``complete'' set of conserved quantities,
the perpendicular component $\hat{A}_{\perp}$ is negligible in the thermodynamic limit.
More precisely, the diagonal component $\hbO_{\perp}$,
which is relevant in the long-time average
(see Eq.~\eqref{eq:diag}),
becomes negligible.
\end{comment}

%##########################################################
%######## 全段落を受けて、熱力学極限で厳密にGGEが一致する場合の議論#########################
%##########################################################
The observable projection idea readily justifies the GGE
in the thermodynamic limit as follows.
Note that the long-time average for the actual dynamics
is
$\braket{\hat{A}}_{\text{LT}}=\braket{\hO_\parallel}_{\text{LT}}+\braket{\hat{A}_{\perp}}_{\text{LT}}
=\braket{\hO_\parallel}_{\text{ini}}+\braket{\hbO_{\perp}}_{\text{ini}}$,
where we have used $\hat{A}_{\parallel}=\sum_{\alpha} p_{A \alpha} \hat{Q}_{\alpha}$
and Eqs.~\eqref{eq:diag} and \eqref{eq:constraints}.
On the other hand, the GGE gives 
$\braket{\hat{A}}_{\text{GGE}}=\braket{\hO_\parallel}_{\text{GGE}}+\braket{\hat{A}_{\perp}}_{\text{GGE}}
=\braket{\hO_\parallel}_{\text{ini}}+\braket{\hbO_{\perp}}_{\text{GGE}}$
since the GGE satisfies $\braket{\hat{Q}_{\alpha}}_{\text{ini}}=\braket{\hat{Q}_{\alpha}}_\text{GGE}$
by definition and then $\braket{\hO_\parallel}_{\text{ini}}=\braket{\hO_\parallel}_{\text{GGE}}$
and $\braket{\dO_{\perp}}_{\text{GGE}}=0$.
Thus the error of the GGE description depends only on the perpendicular component as
\begin{equation}
\braket{\hat{A}}_{\text{LT}}-\braket{\hat{A}}_{\text{GGE}}=\braket{\hbO_\perp}_{\text{ini}}-\braket{\hbO_\perp}_{\text{GGE}},
\end{equation}
which vanishes in the thermodynamic limit if our $\{\hat{Q}_{\alpha}\}$ is complete.
When the set of conserved quantities $\{\hat{Q}_{\alpha}\}$ is incomplete, 
$\hbO_{\perp}$ does not vanish and the GGE prediction deviates from the long-time average in the thermodynamic limit
\hlf{unless $\hbO_\perp$ is accidentally fit by the GGE, i.e., $\braket{\hbO_\perp}_{\text{ini}}=\braket{\hbO_\perp}_{\text{GGE}}$. } 
\hlf{In Appendix~\ref{additionalLIOM},
we show \hlik{that the nonvanishing norm of $\hbO_\perp$ implies}
the existence of the additional local conserved quantities which should be incorporated into $\{\hat{Q}_{\alpha}\}$.}

The above justification of the GGE highlights
the importance of taking enough amount of conserved quantities.
\hlik{In performing the operator projections,
we can also single out relevant conserved quantities for the local observables of interest.
We note that the role of noncommutative conserved quantities has not been clear
in the above discussion,
and it often happens that we only have commutative conserved quantities.}
\hlf{In Appendix~\ref{detection}, we show how to judge whether
noncommutative ones need to be incorporated into the GGE or not.}

%##########################################################
%######## 今回の話と関連付けて、有限サイズでの話に関する言及#########################
%##########################################################
Finite-size systems are also of interest,
in which the long-time average can be influenced
by \hlf{nonlocal}
conserved quantities, which are excluded from the minimal complete set of the conserved quantities \hlf{in the thermodynamic limit}.
Incorporating those conserved quantities, we have smaller errors with GGEs at finite system size or more accurate GGEs.

\hlik{We can apply all the arguments in this section to any system including interacting integrable systems and even nonintegrable systems.
Upon calculating $\hbO$ and its parallel and perpendicular components,
we need to perform numerically
the exact diagonalization of $\hat{H}$ and
inner products between observables and conserved quantities
with their explicit matrix representations.
Thus, the accessible system size is rather limited (see also Sec.~VII for further discussion).
On the other hand, in noninteracting integrable systems, we can do more analytically to get deeper insights, and larger system sizes are accessible.} %the calculations of NCGGE expectation value in larger system size is possible.}  
%\hlf{The calculation of NCGGE expectation value in interacting integrable systems such as XXZ chain  is an interesting open problem.}
In the following, we focus on free fermions
in one dimension and discuss various versions of the NCGGE.

\section{Exact NCGGE at finite system size}
Interestingly, for free fermions in one dimension,
we can analytically construct NCGGE which exactly describes
the long-time average at finite system size.
The construction is step-by-step:
The NCGGE involving all the up-to-$N$-body conserved quantities
exactly describes all the up-to-$N$-body observables.

%##########################################################
%######## \UTF{2464}今回の模型について  #########################
%##########################################################
%
%\setlength{\abovedisplayskip}{2pt} % 上部のマージン
%\setlength{\belowdisplayskip}{2pt} % 下部のマージン
%%%%%セットアップ
We begin by defining the model Hamiltonian
\begin{align}\label{eq:ham}
\hat{H}=- \sum_{i=1}^{L}\left(\hat{c}_{i}^{\dagger} \hat{c}_{i+1}+\mathrm{h.c.}\right)=\sum_k \epsilon_k \hat{c}_{k}^{\dagger} \hat{c}_{k},
\end{align}
where
we have set the transfer integral to unity,
$L$ is the number of sites,
the periodic boundary condition is imposed,
and $\hat{c}_{i}$ ($\hat{c}_{i}^\dag$)
is the annihilation (creation) operator for
the spinless fermion at site $i$:
$\{\hat{c}_{i}, \hat{c}_{j}^{\dagger}\}=\delta_{ij}$
and  $\{\hat{c}_{i},\hat{c}_{j}\}=\{\hat{c}_{i}^{\dagger}, \hat{c}_{j}^{\dagger}\}=0$ for all $i$ and $j$.
We have introduce the Fourier transform
$\hat{c}_{k}=L^{-1/2}\sum_{\hl{j}}e^{-ikj}\hat{c}_{j}$
and $\epsilon_{k}=-2\cos k$,
where $k=2\pi m/L$ $(m\in\mathbb{Z})$
with $-\pi<k\le\pi$.
Thus,
$\sum_{k}$ means the sum over the range
$\{2\pi m /L\ |m=- L/2<m\leq  L/2, m\in \mathbb{Z} \}$.

%##########################################################
%######## 保存量の構成  #########################
%##########################################################
At one-body level,
this Hamiltonian has two kinds of conserved quantities:
\begin{align}\label{eq:cq1}
\I_{k}=\hat{c}_{k}^{\dagger} \hat{c}_{k},\qquad
\J_{k}=\hat{c}_{-k}^{\dagger} \hat{c}_{k}
\end{align}
While only $\I_{k}$ is usually considered in the literature~\cite{Rigol2007},
$\J_{k}$ arising from the double degeneracy \hl{of the dispersion relation in the single-particle spectrum} $\epsilon_{k}=\epsilon_{-k}$ except $k=0$ and $\pi$  \hl{(similarly to the XY chain case in Ref.~\cite{Fagotti_2014}}) is also allowed in the NCGGE.
The set of these conserved quantities are nonconmmutative
due to the algebra 
$[\I_{k},\J_{\pm k}]=\mp\J_{\pm k} $ and $[\J_{k},\J_{-k}]=\I_{-k}-\I_{k}$ (all other commutators vanish) .

\hl{
Note that $\I_{k}$ can be written as the sum of  local conserved quantities (see the supplemental material  of Ref.~\cite{PhysRevLett.124.040603}), but $\J_{k}$ cannot. Taking the Fourier transformation of $\J_{k}$, we have the Wannier-basis form of the additional conserved quantity $\hat{W}_{n}$,
\begin{align}
\hat{W}_{n}
\equiv
\sum_{k}e^{-ink}\J_{k}
=
\sum_{j=1}^{L}\hat{c}^{\dag}_{j+n}\hat{c}_{-j},
\label{nonlocalJ}
\end{align}%
where the site indices $j+n$ and $-j$ should be interpreted in modulo $L$.
\hlb{
Note that
$\hat{W}_{n}$ includes the long-range hopping of $O(L)$ and local hopping with the same weight, which implies that $\J_{k}$ is a nonlocal conserved quantity.}
}

%##########################################################
%######## NCGGEの構成#########################
%##########################################################
We define the GGE with all the one-body
conserved quantities in Eq.~\eqref{eq:cq1}
as the one-body NCGGE:
\begin{align}\label{eq:nc1}
  \hat{\rho}_{\text{1NC}}
  &=
  \frac{1}{Z_{\text{1NC}}}
  \exp\left[- \sum_k (\lambda_k\I_k+\omega_k\J_k)\right],
\end{align}
where
$Z_{\text{1NC}}=\tr e^{- \sum_{k}(\lambda_k\I_k+\omega_k\J_k)}$
and $\lambda_k$ $(=\lambda_k^*)$ and $ \omega_k$ $(=\omega_{-k}^*)$ are the generalized temperatures
determined by
\begin{align}\label{eq:1nc_cond}
	\ave{\I_k}{1NC}=\ave{\I_k}{ini},\qquad
	\ave{\J_k}{1NC}=\ave{\J_k}{ini}
\end{align}
for every $k$.

%##########################################################
%######## 一体物理量が厳密に一致すること#########################
%##########################################################
Remarkably, the one-body NCGGE thus constructed
describe, without an error, long-time averages
of all the one-body observables.
To show this,
we take an arbitrary
one-body observable $\hat{A}^{(1)}=\sum_{k,q}A_{kq}\hat{c}^{\dag}_{k}\hat{c}_{q}$
and consider its long-time average.
Utilizing the Heisenberg picture,
$\hat{c}_{k}(t)\equiv e^{i\hH t}\hat{c}_ke^{-i\hH t}
=e^{-i\epsilon_k t}\hat{c}_k$
and $\hat{c}_{k}^\dag(t)=e^{i\epsilon_k t}\hat{c}_k^\dag$,
we obtain
$\ave{\hat{A}^{(1)}}{LT}
=\sum_{k,q}A_{kq}\ave{\hat{c}^{\dag}_{k}\hat{c}_{q}}{ini}
\overline{e^{i(\epsilon_k-\epsilon_q)t}}
=\sum_{k}(A_{kk}\braket{\I_{k}}_{\text{ini}}+A_{k,-k}\braket{\J_{k}}_{\text{ini}})$.
We emphasize that the long-time average has been nonvanishing only
for $\epsilon_k=\epsilon_q$
and this condition is equivalent to
that $\hat{c}^{\dag}_{k}\hat{c}_{q}$ is a conserved quantity
since $[\hH,\hat{c}^{\dag}_{k}\hat{c}_{q}]=(\epsilon_k-\epsilon_q)\hat{c}^{\dag}_{k}\hat{c}_{q}$.
On the other hand,
we have, for the one-body NCGGE,
$\ave{\hat{A}^{(1)}}{1NC}
=\sum_{k}(A_{kk}\braket{\I_{k}}_{\text{ini}}+A_{k,-k}\braket{\J_{k}}_{\text{ini}})$
since $\ave{\hat{c}^{\dag}_{k}\hat{c}_{q}}{1NC}=0$
for $\epsilon_k\neq\epsilon_q$.
Using Eq.~\eqref{eq:1nc_cond},
we obtain
\begin{align}\label{eq:1nc_A}
\ave{\hat{A}^{(1)}}{1NC}=\ave{\hat{A}^{(1)}}{LT}\qquad \forall \hat{A}^{(1)}
\end{align}
even when the system size $L$ is finite.
This is a remarkable property
that the conventional CGGE does not have.
%\hl{We also note that this argument is applied to both local observables and nonlocal observables.}
The CGGE density matrix $\hat{\rho}_{\text{C}}$ is defined only with $\I_k$ 
and cannot be exact at finite $L$,
$\ave{\hat{A}^{(1)}}{C}
=\sum_{k}A_{kk}\braket{\I_{k}}_{\text{ini}}\neq\ave{\hat{A}^{(1)}}{LT}$.

%##########################################################
%######## それの2体物理量バージョン#########################
%##########################################################
The above exactness of the one-body NCGGE
naturally let us find the exact $N$-body NCGGE.
Let us first consider the $N=2$ case
and take a two-body observable
$\hat{A}^{(2)}=\sum_{k_1,k_2,q_1,q_2}A_{k_{1}k_{2};q_{1}q_{2}}\hat{c}^{\dag}_{k_{1}}\hat{c}^{\dag}_{k_{2}}\hat{c}_{q_{2}}\hat{c}_{q_{1}}$.
Its long-time average is given by
$\ave{\hat{A}^{(2)}}{LT}
=\sum_{k_1,k_2,q_1,q_2}'A_{k_{1}k_{2};q_{1}q_{2}}
\braket{ \hat{c}^{\dag}_{k_{1}}\hat{c}^{\dag}_{k_{2}}\hat{c}_{q_{2}}\hat{c}_{q_{1}}}_{\text{ini}}$,
where $\sum'$ means the restriction of the sum
to $\epsilon_{k_1}+\epsilon_{k_2}=\epsilon_{q_1}+\epsilon_{q_2}$.
Here we note that every $\hC_{k_1k_2q_1q_2}\equiv\hat{c}^{\dag}_{k_{1}}\hat{c}^{\dag}_{k_{2}}\hat{c}_{q_{2}}\hat{c}_{q_{1}}$
in the restricted sum is a conserved quantity.
%二体保存量の紹介
These two-body conserved quantities include the products of two one-body conserved quantities~\eqref{eq:cq1}
%, i.e., $\I_{k}\I_{q}, \I_{k}\J_{q}, \J_{k}\J_{q}$. 
%偶然縮退があることをいう
as well as others
due to accidental degeneracy such as $c^{\dag}_{\frac{\pi}{2}-k}c^{\dag}_{\frac{\pi}{2}-q}c_{\frac{\pi}{2}+q}c_{\frac{\pi}{2}+k}$.
If we define the \hlf{exact} two-body NCGGE $\hatrho_\text{e2NC}$
by $\ave{\hC_{k_1k_2q_1q_2}}{e2NC}=\ave{\hC_{k_1k_2q_1q_2}}{ini}$
and Eq.~\eqref{eq:1nc_cond} for e2NC,
one can easily show 
$\ave{\hat{A}^{(n)}}{e2NC}=\ave{\hat{A}^{(n)}}{LT}\ (\forall \hat{A}^{(n)})$
for ($n=1$ and 2).
Thus, we have obtained the NCGGE
that describes the long-time average of 
each one- or two-body observable
exactly at finite $L$.
In a similar manner,
we can systematically construct the \hlf{exact} $N$-body GGE
that is exact for all up-to-$N$-body observables
at finite system size.
\hl{
Note that the conserved quantities used in the $N$-body NCGGE are non-local except for $\I_{k}$.
}

%######
In practice,
it is a hard task
both analytically and numerically
to determine all the generalized temperatures
for the exact $N$-body NCGGE for $N\ge2$
since it is essentially a many-body problem.
\hl{However, it is conceptually important:
\hlb{There exists a systematic construction of the GGE that
is exact for \emph{all} the less-than-$N$-body observables
at \emph{finite} system size.
}}
Below, we discuss some special NCGGEs of practical relevance:
the exact one-body and approximate two-body NCGGEs.

%######################################
\section{Application of exact one-body NCGGE}
As shown above,
the one-body NCGGE~\eqref{eq:nc1}
exactly describes all the one-body observables
unlike the conventional CGGE.
We further study how this NCGGE works
for two-body observables.
Fortunately,
we can analytically obtain the generalized temperatures
$\lambda_{k}$ and $\omega_k$.
Although we leave the detail
in Appendix~\ref{generalized temperature},
an important idea is
to perform a unitary transformation
in each $(k,-k)$ subspace:
$(\hat{d}_{k}^{\dag},\hat{d}_{-k}^{\dag})=(\hat{c}_{k}^{\dag},\hat{c}_{-k}^{\dag})U_k$,
which diagonalizes the exponent in Eq.~\eqref{eq:nc1}.
Then we have a diagonal form
\begin{align}\label{eq:nc1diag}
\hat{\rho}_{\text{1NC}}
=
\frac{1}{Z_{\text{1NC}}}
 \prod_{k}
  \exp\left(-\eta_k \Id_{k}\right),
\end{align}%
where $\Id_{k}=\hat{d}^{\dag}_{k}\hat{d}_{k}$ is the conserved quantity
in the new basis and $\eta_{k}$ is some linear combination
of $\lambda_{k}$ and $\omega_k$.
Equation~\eqref{eq:nc1diag} is useful
for obtaining the generalized temperatures
(see Appendix~\ref{generalized temperature}).

\begin{figure}[tb]
 \centering
 \includegraphics[width=\columnwidth]{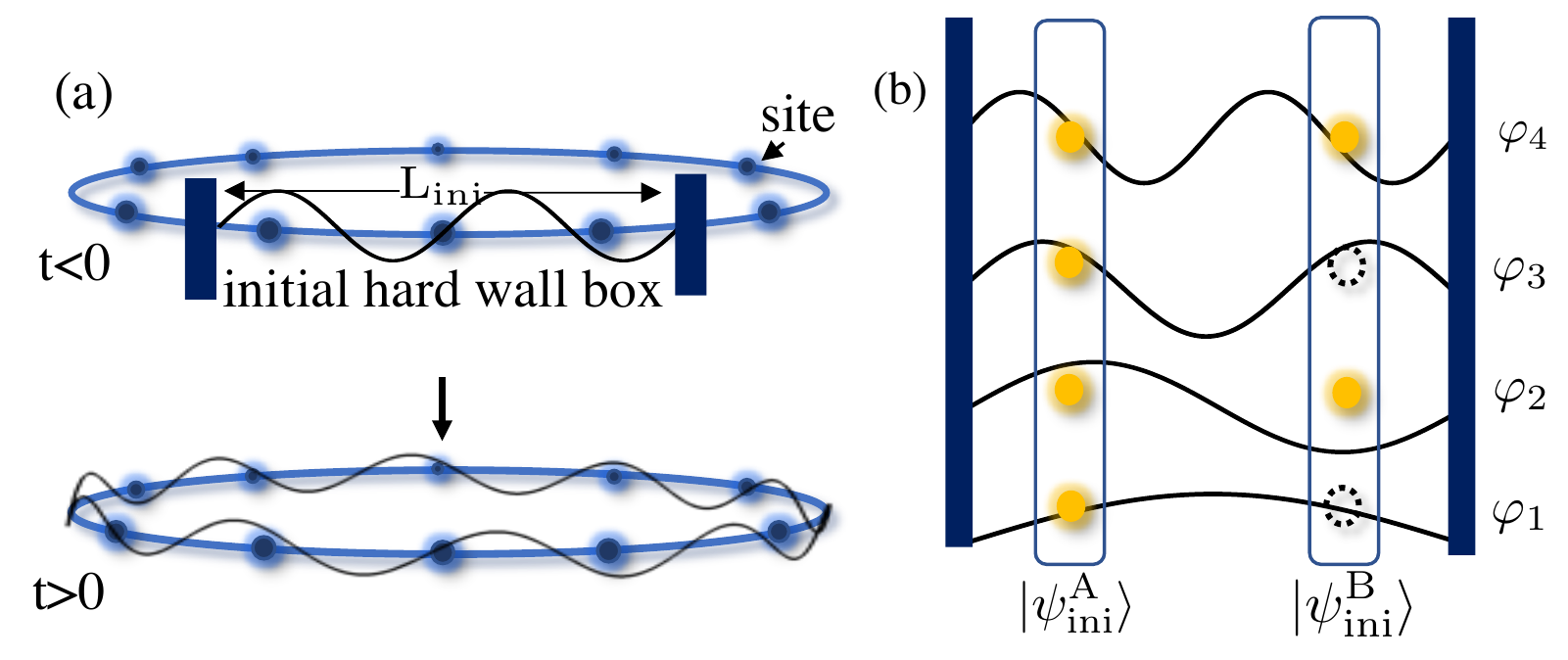}
\caption{(a) Schematic illustration of dynamics protocol.
(b) Illustration of two initial states $\ket{\psi_{\text{ini}}^{\text{A}}}$
and $\ket{\psi_{\text{ini}}^{\text{B}}}$.
Filled circles represent the occupied one-particle
energy eigenstates.
}
\label{initial}
\end{figure}

%#####
To test the accuracy of $\hatrho_\text{1NC}$,
we consider a concrete initial state and
its dynamics under the Hamiltonian~\eqref{eq:ham}.
As shown in Fig.~\ref{initial}(a),
we suppose an initial hard wall box,
which confines $N$ particles to the sites $1\le i\le\Lin$
($N\le\Lin$).
The one-particle energy eigenstates within the box are 
$\varphi_{n}(j)=(\Lin+1)^{-1/2}\sin [\pi n j/(\Lin+1)]$
as illustrated in Fig.~\ref{initial}(b).
Introducing the creation operators for these eigenstates
as $\Phi_n^\dag=\sum_{j=1}^{\hlf{L_\text{ini}}}
\varphi_{n}(j)
\hat{c}^{\dag}_{j}$,
we consider the following two initial states:
the ground state $\ket{\psi_{\text{ini}}^{\text{A}}}
=\prod_{n=1}^{N}\Phi_n^\dag\ket{0}$
and an excited state
$\ket{\psi_{\text{ini}}^{\text{B}}}
=\prod_{n=1}^{N}\Phi_{2n}^\dag\ket{0}$
(for $N\le\Lin/2$).
We remove the hard wall instantaneously at time $t=0$,
let these initial states evolve under $\hH$, or freely expand
into the entire $L$ sites,
and analyze the long-time average of various observables.

\begin{figure}[tb]
 \centering
 \includegraphics[width=\columnwidth]{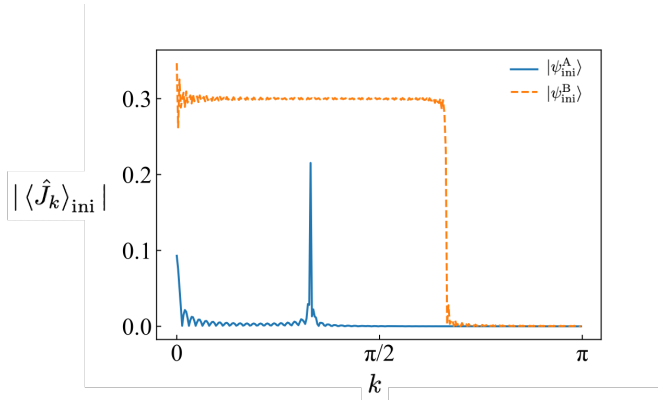}
\caption{
\hl{
The values of  additional conserved quantities $|\braket{\J_{k}}_{\text{ini}}|$ in the initial states $\ket{\psi^{\text{A}}_{\text{ini}}}$ and $\ket{\psi^{\text{B}}_{\text{ini}}}$ with $L=600$, $L_{\text{ini}}=360$, and $N=120$.
Results
are not shown for $k < 0$ since $|\braket{\J_{k}}_{\text{ini}}|=|\braket{\J_{-k}}_{\text{ini}}|$ .
}
}
\label{iom}
\end{figure}

\hl{
Figure~\ref{iom} displays the values of the additional conserved quantities
$|\braket{\J_{k}}_{\text{ini}}|\equiv |\bra{\psi_{\text{ini}}}\J_{k}\ket{\psi_{\text{ini}}}|$ for $\ket{\psi_{\text{ini}}}=\ket{\psi^{\text{A}}_{\text{ini}}}$ and $\ket{\psi^{\text{B}}_{\text{ini}}}$.
%Note that $\J_{k}$ is traceless and the origin of the spectrum of $\J_{k}$ is 0.
\hlb{While $|\braket{\J_{k}}_{\text{ini}}|$ are almost zero for most $k$ in $\ket{\psi^{\text{A}}_{\text{ini}}}$,
it has large values for $0<k<2\pi/3$ in $\ket{\psi^{\text{B}}_{\text{ini}}}$.
Thus, $\J_{k}$ is less important for the GGE in case of $\ket{\psi^{\text{A}}_{\text{ini}}}$ and the generalized temperatures for $\J_{k}$ are almost zero for most $k$. On the other hand, in case of $\ket{\psi^{\text{B}}_{\text{ini}}}$, we should use $\J_{k}$ in the GGE and the generalized temperatures have large absolute values.}
%From now on we concentrate on the initial state $\ket{\psi^{\text{B}}_{\text{ini}}}$.
}
%## schematic

%### fig comparision
\begin{figure}[tb]
\centering
\includegraphics[width=\columnwidth]{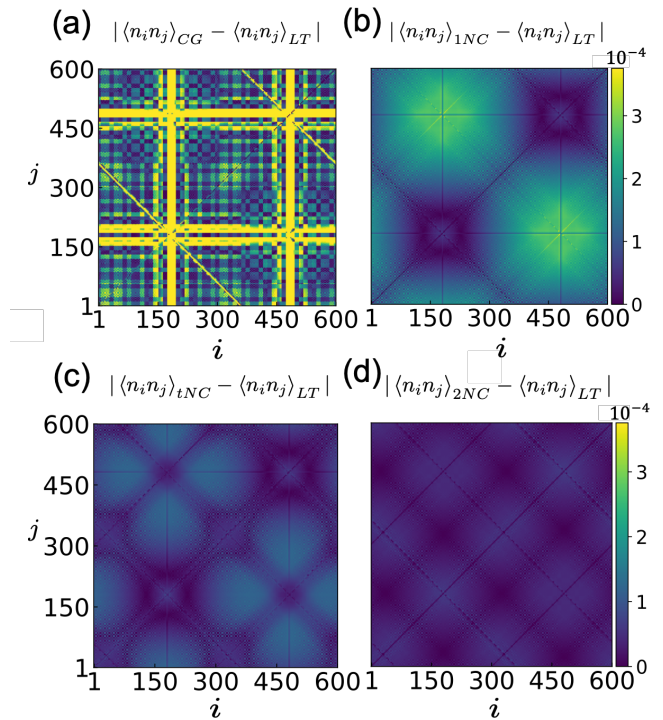}
\caption{
\hl{
Error of GGEs $|\braket{\hat{n}_{i}\hat{n}_{j}}_{\mathrm{GGE}}-\braket{\hat{n}_{i}\hat{n}_{j}}_{\mathrm{LT}}|$
for the density-density correlation between sites $i$ and $j$
calculated with the (a) CGGE, (b) one-body NCGGE,  (c) trigonal NCGGE, and (d) two-body NCGGE with $L=600$, $L_{\text{ini}}=360$, and $N=120$.
In all panels, we use the initial state $\ket{\psi_{\text{ini}}^{\text{B}}}$,
and implicitly assume the normal ordering for $\hat{n}_{i}\hat{n}_{j}$ (see footnote [57]).
}
}
\label{error}
\end{figure}

\begin{figure}[tb]
 \centering
 \includegraphics[width=\columnwidth]{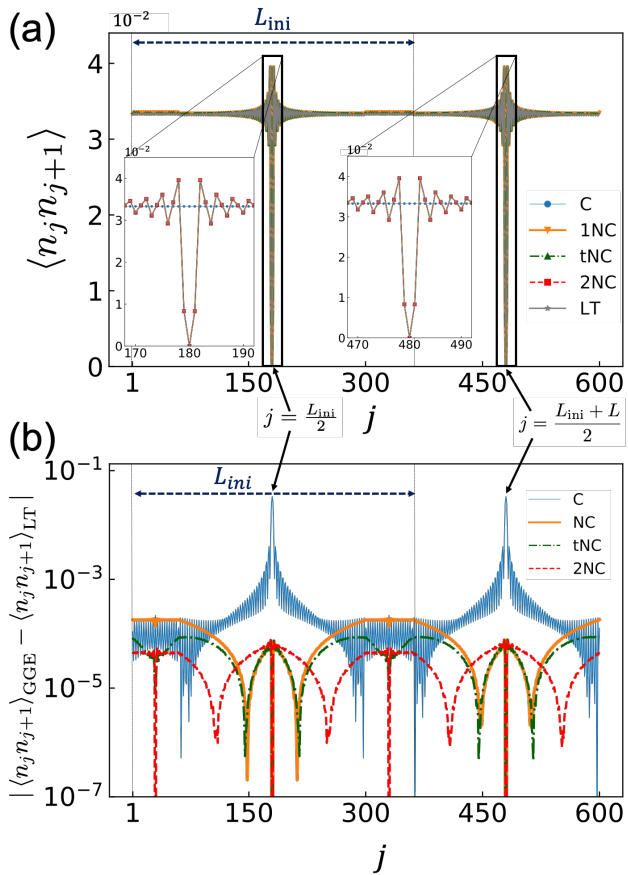}
\caption{\hl{
(a) Expectation values of local density-density correlation of $\braket{n_jn_{j+1}}$ in the CGGE, one-body NCGGE, trigonal NCGGE, two-body NCGGE, and long-time average for the initial state $\ket{\psi_{\text{ini}}^{\text{B}}}$ with $L=600$, $L_{\text{ini}}=360$, and $N=120$. There are the characteristic peaks which cannot be captured by the CGGE  at the high-symmetry points of the initial state $\ket{\psi_{\text{ini}}^{\text{B}}}$, $j=L_{\text{ini}}/2$ and $(L_{\text{ini}}+L)/2$. 
At the high-symmetry points, the expectation value of the correlation function in long-time average and NCGGEs are zero, but the CGGE does not.
(b) Error of GGEs $|\braket{\hat{n}_{j}\hat{n}_{j+1}}_{\mathrm{GGE}}-\braket{\hat{n}_{j}\hat{n}_{j+1}}_{\mathrm{LT}}|$ for the local density-density correlation between sites $j$ and $j+1$.
}
}
\label{density}
\end{figure}

\begin{figure}[tb]
 \centering
 \includegraphics[width=\columnwidth]{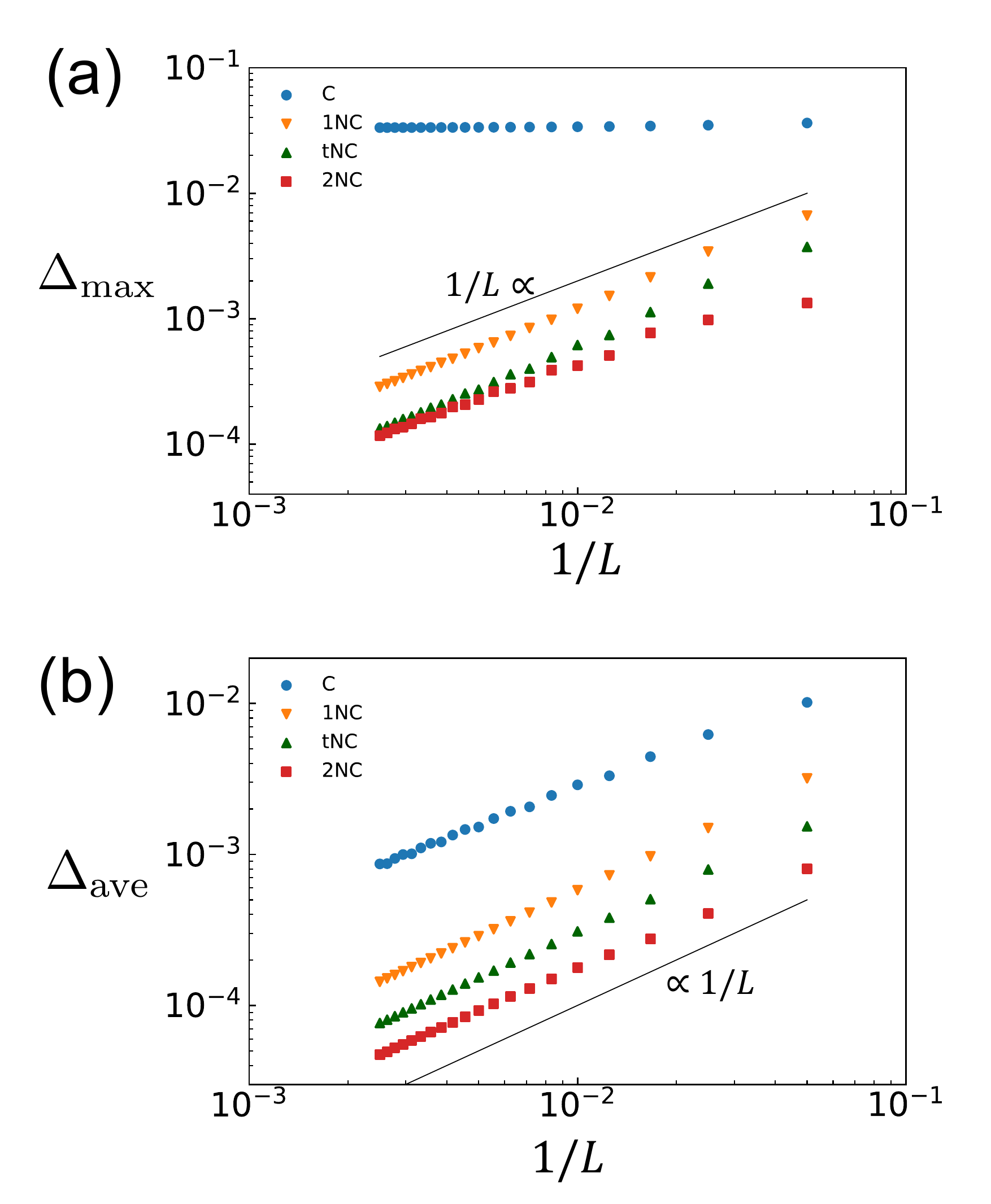}
\caption{
\hl{
The $L$-dependence of (a) maximum  $\Delta_{\text{max}}$ errors and (b) averaged  $\Delta_{\text{ave}}$ errors of the expectation values of local density-density correlation of  $\braket{n_jn_{j+1}}$ in the GGEs calculated with $\Lin/L=3/5$ and $N/\Lin=1/3$ held fixed.
\hlb{The initial state is $\ket{\psi_{\text{ini}}^{\text{B}}}$, and $L$, $L_{\text{ini}}$, and $N$ are all even
at every data point.}
}
}
\label{lplot}
\end{figure}

%#######
To compare the one-body NCGGE and the conventional CGGE,
we consider some two-body observables
since we have already shown that
one-body observables are exactly described
by the one-body NCGGE.
%For a clear comparison,
\hlb{To highlight the role of $\J_{k}$,}
we take $\ket{\psi_{\text{ini}}^{\text{B}}}$
\hl{as an initial state}
and focus on the density-density
correlation $\hat{n}_{i}\hat{n}_{j}$
($n_{i}\equiv\hat{c}^{\dag}_{i}\hat{c}_{i}$)~\footnote{
We actually calculate the normal-ordered operator $:\!\hat{n}_{i}\hat{n}_{j}\!:\,=\hat{c}_i^\dag\hat{c}_j^\dag\hat{c}_j\hat{c}_i$
to remove the unwanted one-body contributions.
}
and
calculate the error of the GGEs $|\braket{\hat{n}_{i}\hat{n}_{j}}_{\mathrm{GGE}}-\braket{\hat{n}_{i}\hat{n}_{j}}_{\mathrm{LT}}|$, where $\mathrm{GGE}$ means the one-body NCGGE (1NC) or CGGE (C).
We plot these errors in Figs.~\ref{error}(a) and (b),
finding $\hat{\rho}_{\text{1NC}}$ more accurate than the CGGE as a whole.
\hl{
We turn our attention further to local physical quantities $n_{j}n_{j+1}$, which are 1-local operators and the  sub-diagonal components of Fig.~\ref{error}.
For a quantitative comparison of the local observables,
we plot the expectation values of $\hat{n}_{j}\hat{n}_{j+1}$ in Fig.~\ref{density}(a) and the errors of the GGEs $|\braket{\hat{n}_{j}\hat{n}_{j+1}}_{\mathrm{GGE}}-\braket{\hat{n}_{j}\hat{n}_{j+1}}_{\mathrm{LT}}|$ in Fig.~\ref{density}(b).
}
We find that $\hat{\rho}_{\text{1NC}}$ describes
the long-time average $\braket{\hat{n}_{j}\hat{n}_{j+1}}_\text{LT}$ better than the CGGE for most $j$ in Fig.~\ref{density}.
It is noteworthy that
the $\hat{\rho}_{\text{1NC}}$
captures the characteristic peaks of $\braket{\hat{n}_{j}\hat{n}_{j+1}}_\text{LT}$ while the CGGE cannot.
These characteristic peaks are related to the inversion symmetry
and not present for $\ket{\psi_{\text{ini}}^{\text{A}}}$ (see Appendix~\ref{ground initial}),
for which the improvement by $\hat{\rho}_{\text{1NC}}$.

%#####
We also examine how the errors scale in the system size $L$
with ratios $N/\Lin$ and $\Lin/L$ held fixed.
We define the averaged error of the density-density correlation
by $\Delta_\text{ave}\equiv\sum_{j}|\braket{\hat{n}_{j}\hat{n}_{j+1}}_{\mathrm{GGE}}-\braket{\hat{n}_{j}\hat{n}_{j+1}}_{\mathrm{LT}}|/L$,
which is plotted for GGEs
at several system sizes in Fig.~\ref{lplot}(b).
The error is much smaller for $\hat{\rho}_{\text{1NC}}$,
and decreases as $\propto1/L$
to vanish in the thermodynamic limit for both GGEs~\footnote{This power-law decay is a feature of integrable models~\cite{Biroli2010}, and the error decays exponentially in nonintegrable models~\cite{Beugeling2014,Ikeda2015}}.
Thus, the CGGE also becomes accurate in this limit on average.
However, when we use a more strict definition for the error
defined by $\Delta_\text{max}\equiv\max_{j}|\braket{\hat{n}_{i}\hat{n}_{j+1}}_{\mathrm{GGE}}-\braket{\hat{n}_{j}\hat{n}_{j+1}}_{\mathrm{LT}}|$,
we come to a different conclusion:
the one-body NCGGE becomes accurate as $L\to\infty$
while the CGGE does not, as shown in Fig.~\ref{lplot}(a).
This is due to the characteristic peaks shown in Fig.~\ref{density} and 
\hl{the maximum error \hlb{of the CGGE occurs at the high-symmetry points $j=L_{\text{ini}}/2$ and $ (L_{\text{ini}}+L)/2$.}}
\hl{For \hlb{the other} initial state
$\ket{\psi_{\text{ini}}^{\text{A}}}$, as $L$ increases,
 $\Delta_{\text{max}}$ of the CGGE also decreases as $\propto 1/L$ because there are no characteristic peaks, which cannot be captured by the CGGE.  }
These results show that
\hl{the one-body NCGGE improves the GGE prediction quantitatively as a whole, but some of the local correlations such as $n_{L_{\text{ini}}/2}n_{L_{\text{ini}}/2+1}$ can be improved by one-body NCGGE  qualitatively from the CGGE.}
The NCGGE can be necessary
for accurately describing the actual stationary state
even in the thermodynamic limit, depending on the initial state.

%######################
\section{Improvement of exact one-body NCGGE}
%######################
\subsection{Trigonal NCGGE}
Although it is difficult to implement the exact two-body NCGGE,
we can partly include two-body conserved quantities,
improving the one-body NCGGE.
To inspect which conserved quantities are important,
we calculate  $|\braket{\Id_{k}\Id_{q}}_{\text{ini}}-\braket{\Id_{k}\Id_{q}}_\text{1NC}|$,
and find  that most deviations reside around
the diagonal ($k=q$) and anti-diagonal ($k=-q$) components (see Appendix~\ref{2fit}).
Noting that $(\Id_k)^{2}=\Id_k$, we take the products of the adjacent pairs
$\Id_k\Id_{k+\Delta k}$
with $\Delta k=2\pi/L$,
defining the following trigonal NCGGE:
%密度行列を提示
\begin{align}
 \hat{\rho}_{\text{tNC}}
  =
  \frac{1}{Z_{\text{tNC}}}
  \exp\left[- \sum_{k}\left(\tilde{\eta}_k\Id_k+\Lambda_{k}\Id_k\Id_{k+\Delta k}\right)\right],
 \end{align}%
where $Z_{\text{tNC}}$ is defined by $\tr \hat{\rho}_{\text{tNC}}=1$.
Remarkably,
we can efficiently obtain the generalized temperatures $\tilde{\eta}_k$ and $\Lambda_{k}$
numerically
by a method similar to the transfer matrix
for the one-dimensional Ising model (see Appendix~\ref{tnc temp}).

%#####
The trigonal NCGGE thus implemented
leads to a quantitative improvement
of the one-body NCGGE.
The error of the two-body conserved quantities $\Id_{k}\Id_{q}$ in both initial state $\ket{\psi_{\text{A}}}$ and $\ket{\psi_{\text{B}}}$
is reduced near the diagonal ($k=q$) components (see Appendix~\ref{2fit}).

\subsection{two-body NCGGE}
When we take all the two-body conserved quantities $\Id_{k}\Id_{q}$ into the GGE, the explicit calculation of the generalized temperatures is a very hard task.
We call this ideal ensemble as the two-body NCGGE. 
We remark that this two-body NCGGE is different from
the exact two-body NCGGE, which also involves two-body conserved quantities not in the form of $\Id_{k}\Id_{q}$.

Interestingly, without having the generalized temperatures,
we can calculate the expectation value of the observables in the
two-body NCGGE in the free fermion model from the information of the initial conditions.
The density matrix of two-body NCGGE  is  formally written as 
\begin{align}
 \hat{\rho}_{\text{2NC}}
  &=
  \frac{1}{Z_{\text{2NC}}}
  \exp\left(- \sum_{k}\tilde{\eta}_k\Id_k-\sum_{k>q}\Lambda_{kq}\Id_k\Id_q\right),
  \end{align}%
  \hlf{where $Z_{\text{2NC}}$ is defined by $\tr \hat{\rho}_{\text{2NC}}=1$.}
%We cannot calculate explicitly $\tilde{\eta}_{k}$ and $\Lambda_{kq}$ unlike trigonal NCGGE or one-body NCGGE.
%However, we can calculate the expectation value of
Let us consider a general two-body observable $\hat{A}^{(2)}=\sum_{k_1,k_2,q_1,q_2}\tilde{A}_{k_{1}k_{2};q_{1}q_{2}}\hat{d}^{\dag}_{k_{1}}\hat{d}^{\dag}_{k_{2}}\hat{d}_{q_{2}}\hat{d}_{q_{1}}$ in the two-body NCGGE.
In taking its expectation value for $\hat{\rho}_{\text{2NC}}$,
only two kinds of contributions $k_1=q_1$ and $k_2=q_2$ or $k_1=q_2$ and $k_2=q_1$
are nonvanishing
\begin{align}
\ave{\hat{A}^{(2)}}{2NC}
&=\sum_{kq}(\tilde{A}_{kq;kq}-\tilde{A}_{kq;qk})\braket{\Id_{k}\Id_{q}}_{\text{2NC}}\\
&=\sum_{kq}(\tilde{A}_{kq;kq}-\tilde{A}_{kq;qk})\braket{\Id_{k}\Id_{q}}_{\text{ini}}.
\end{align}
To obtain the last equality, we have used the determining equations
for the generalized temperatures, $\braket{\Id_{k}\Id_{q}}_{\text{2NC}}=\braket{\Id_{k}\Id_{q}}_{\text{ini}}$.

%\hl{
%Note that the two-body conserved quantities used in the two-body NCGGE and the trigonal NCGGE are  non-local conserved quantities.
%}

\hl{
We plot in Fig.~\ref{error}
the error of the trigonal NCGGE (c) and the two-body NCGGE (d)
for the density-density correlation
$\hat{n}_{j}\hat{n}_{j+1}$, where  the initial state  is $\ket{\psi_{\text{ini}}^{B}}$.
In Fig.~\ref{lplot}, we observe qualitative features including $\Delta_\text{ave}\propto1/L$
similar to those of the one-body NCGGE.
The more two-body nonlocal conserved quantities we take into the GGE, the more the GGE predictions of the non-local correlations are improved (which corresponds to the much-off-diagonal element in Fig.~\ref{error}). 
We can see significant reductions of the errors $\Delta_\text{ave}$ and $\Delta_\text{max}$ in Fig.~\ref{lplot}
in both the trigonal NCGGE and the two-body NCGGE and the reductions are larger in  the two-body NCGGE than in  the trigonal NCGGE.
}

%##########################################################
%######## Summary
%#######################################################
\section{Summary and Outlook}
Introducing noncommutative sets of conserved quantities
and the observable projection idea,
we have systematically shown that the NCGGE describe the long-time behavior
of isolated quantum systems better than the conventional CGGE.
For noninteracting integrable systems,
we have explicitly constructed the exact $N$-body NCGGE that
describes the long-time average of up-to-$N$-body observables
without an error even at finite system size.
Besides, we have shown that the one-body NCGGE,  the trigonal NCGGE, and the two-body NCGGE can be numerically implemented
and describe two-body observables well.
\hl{
We note that the additional noncommutative conserved quantities are nonlocal.
\hlb{However, there exist local observables which need these} nonlocal additional conserved quantities for qualitative description depending on the initial state.
}

\hlik{
The implementation of the NCGGE
%to other systems such as
in interacting integrable systems
is an important open problem.
%The NCGGE may resolve some known failures of the conventional CGGE.
This problem is challenging 
%The calculation of the NCGGE expectation value  in interacting integrable systems is interesting but  very difficult at present,
because %we do not even know whether there is
noncommutative sets of conserved quantities are not explored well in these systems
and our approach presented for noninteracting systems does not apply. 
However, as we have remarked in Sec.~III,
one can numerically conduct the \hlf{observable projection} scheme for moderate system sizes
and test whether the NCGGE is necessary or not.
The numerical operator projection can be performed
by the Gram--Schmidt orthogonalization \hlf{and by the exact simultaneous diagonalization of the commutative local conserved quantities }
once we know the explicit form of the commutative local conserved quantities 
%$\hat{Q}_{\alpha}^{\text{C}}$
derived from the transfer matrix \hlf{(see Appendix~\ref{detection} for more details)}.
\hlf{With this analysis, we can see whether NCGGE is needed or not in interacting integrable systems. 
 Moreover, we may even find new conserved quantities that do not commute with the well-known commutative ones.}
%In interacting integrable systems, we can calculate  $\hbO\ \!\!^{\text{NC}}$ numerically by exact diagonalization in small system size  if we know the explicit form of the local  conserved quantities $\hat{Q}_{\alpha}^{\text{C}}$ obtained from the expansion of  transfer matrix in terms of a spectral parameter and Gram Schmidt orthogonalization. 
%Note that the explicit form of all the local conserved quantities in the XYZ chain is discovered very recently by some of the authors \cite[] and can be applied to this argument.
%We can test whether the NCGGE is necessary or not by extrapolation of  $\|\hbO\ \!\!^{\text{NC}}\|$ and $\braket{\hbO\ \!\!^{\text{NC}}}_{\text{ini}}$ in the thermodynamic limit.
}

\hlik{Finally,}
we remark that the quantum-information-theoretic thermodynamics using noncommutative conserved quantities
has attracted attention~\cite{YungerHalpern2016}.
An experimental protocol for its realization
has been proposed in small nonintegrable systems~\cite{PhysRevE.101.042117}.
%a system where the small subsystem thermalizes to the thermal state written by noncommutative conserved quantities (this is different from our NCGGE) was proposed using non-integrable systems
%although the discussion of when and how the NCGGE appears has not been deepened.
Our NCGGE 
arising in large integrable systems
provides another route to 
the quantum-information-theoretic thermodynamics.

%#######
\section{Acknowledgements}
We thank H. Tsunetsugu for fruitful discussion and L. Piroli for his helpful comment on our manuscript.
This work was supported by JSPS KAKENHI Grant No.~JP18K13495.
\hlf{K.F. acknowledges support by the Forefront Physics and Mathematics Program to Drive Transformation (FoPM), WINGS Program, the University of Tokyo.}

\appendix

\section{The uniqueness of the generalized temperature in NCGGE}\label{uniqueness}
We will show the generalized temperature of the NCGGE can be determined uniquely. In other words, we will show the equation $\braket{\hat{Q}_{\alpha}}_{\text{ini}}=\tr(\hatrho_{\mathrm{GGE}}\hat{Q}_\alpha)$ has a unique solution for $\{\lambda_{\alpha}\}$ if the conserved quantities are linearly independent.
Let $\mathcal{S}$ denote  the real linear space spanned by the linearly independent set of conserved quantities $\{\hat{Q}_{\alpha}\}$. \hlm{All the elements of $\mathcal{S}$ are hermitian conserved quantities.}
The exponent of the GGE $\hat{X}\equiv -\sum_{\alpha}\lambda_{\alpha}\hat{Q}_{\alpha}$ \hlm{belongs to} $\mathcal{S}$, and $\hatrho_{\text{GGE}}$ is written as \hlm{$\hatrho_{\text{GGE}}=e^{\hat{X}}/\tr e^{\hat{X}}$}.
Substituting  \hlm{$\hatrho_{\text{GGE}}=e^{\hat{X}}/\tr e^{\hat{X}}$} into the entropy $\Psi$ \hlm{in Eq.~\eqref{eq:psi}}, we get
\begin{align}
\Phi(\hat{X})\equiv\Psi(\hatrho_{\text{GGE}},\{\lambda_{\alpha}\})=\log{\tr e^{\hat{X}}}-\braket{\hat{X}}_{\text{ini}}.
\end{align}%
The problem is reduced to the proof of the convexity of $\Phi(\hat{X})$ over $\mathcal{S}$, more specifically, the  proof of the inequality $\Phi((\hat{X}_{1}+\hat{X}_{2})/2)\leq (\Phi(\hat{X}_{1})+\Phi(\hat{X}_{2}))/2$, where $\hat{X}_{1}, \hat{X}_{2} \in \mathcal{S}$ are arbitrary. The second terms of $\Phi$ are canceled, and what we should prove becomes
\begin{align}
\left(\tr e^{\frac{1}{2}(\hat{X}_{1}+\hat{X}_{2})}\right)^{2}
\leq 
\left(\tr e^{\hat{X}_{1}}\right)\left(\tr e^{\hat{X}_{2}}\right).\label{ineq}
\end{align}%
When $\{\hat{Q}_{\alpha}\}$ is \hlm{a} commutative set,  (\ref{ineq}) immediately holds by using the Cauchy-Schwarz inequality \hlm{as discussed below in the noncommutative case.}

\hlm{When $\{\hat{Q}_{\alpha}\}$ is \hlm{a} noncommutative set,} we can utilize the Golden-Thompson inequality \cite{PhysRev.137.B1127} to the left hand side of (\ref{ineq}) because $e^{\frac{1}{2}\hat{X}_{1}}$ and $e^{\frac{1}{2}\hat{X}_{2}}$ are positive semi-definite. Then we can see 
\begin{align}
\tr e^{\frac{1}{2}(\hat{X}_{1}+\hat{X}_{2})}
\leq
\tr e^{\frac{1}{2}\hat{X}_{1}}e^{\frac{1}{2}\hat{X}_{2}}.\label{golden}
\end{align}
Calculating the trace of rhs of (\ref{golden}) with \hlm{an} arbitrary basis $\{\ket{i}\}$ and using the Cauchy-Schwarz inequality, we find
\begin{align}
\left(\tr e^{\frac{1}{2}\hat{X}_{1}}e^{\frac{1}{2}\hat{X}_{2}}\right)^{2}
&=\left(\sum_{i, j} \braket{i|e^{\frac{1}{2}\hat{X}_{1}}|j}\braket{j|e^{\frac{1}{2}\hat{X}_{2}}|i}\right)^{2}
\nonumber\\
&\leq
\left(\sum_{i, j} |\braket{i|e^{\frac{1}{2}\hat{X}_{1}}|j}|^{2}\right)
\left(\sum_{i, j} |\braket{j|e^{\frac{1}{2}\hat{X}_{2}}|i}|^{2}\right)
\nonumber\\
&\ \ (\because \text{Cauchy-Schwarz inequality})
\nonumber\\
&=
\left(\sum_{i} \braket{i|e^{\hat{X}_{1}}|i}\right)
\left(\sum_{i} \braket{i|e^{\hat{X}_{2}}|i}\right)
\nonumber\\
&=
\left(\tr e^{\hat{X}_{1}}\right)\left(\tr e^{\hat{X}_{2}}\right).
\end{align}%
The \hlm{equality condition} of the inequality	is $\hat{X}_{1}\propto \hat{X}_{2} $, which \hlm{does not hold in the} case that $\hat{X}_{1}$ and $\hat{X}_{2} $ do not commute.
This completes the proof. Then we can see $\Phi(\hat{X})$ is convex \hlm{over} $\mathcal{S}$ and there is the unique minimum $\hat{X}^{\ast}$. \hlm{Since} $\{\hat{Q}_{\alpha}\}$ is independent each other, the coefficients of  $\hat{X}^{\ast}$ are determined uniquely, and these coefficients are the unique solution of the generalized temperatures. Note that this proof is the natural extension of the commutative case.

\begin{comment}
\section{The condition for the existence of the additional local conserved quantities}\label{additionalLIOM}
\hlf{
We consider  the condition for the existence of the additional local conserved quantities which should be incorporated into the GGE from Ref.~\cite{PhysRevLett.124.040603}. When  $\hat{A}$ is normalized as $0<\lim _{L \rightarrow \infty}\|\hat{A}\|<\infty$ and  $\lim_{L\rightarrow\infty}\|\hbO_{\perp}\|>0$ where $\|\hat{A}\|\equiv \sqrt{\braket{\hat{A},\hat{A}}}$ , the GGE fails unless the accidental fit by the GGE and  there is an additional local conserved quantities which should be incorporated into the GGE. This is because we can show $\hbO_{\perp}$ has the components of the local conserved quantities. Note that the opposite claim is not true  (see the supplementary material of ~\cite{PhysRevLett.124.040603}). Even when  $\lim_{L\rightarrow\infty}\|\hbO_{\perp}\|=0$, we cannot claim $\hbO_{\perp}$ vanish in the thermodynamic limit as an operator, so we cannot judge there are additional local conserved quantities which should be incorporated into the GGE. 
%The qualitative sufficient condition for the GGE to describe the long-time average is future problem.
}
\end{comment}

\section{Condition for the existence of the additional local conserved quantities}\label{additionalLIOM}
We show a condition for the existence of the additional local conserved quantities which should be incorporated into the GGE. 
 The definition of the complete set of the conserved quantities is the set of all the local or quasilocal conserved quantities. 
 %We restrict the observables in  translationally invariant  and local observables.
The conserved quantity  $\hat{Q}_{\alpha}$ is local when $\langle\hat{A},\hat{Q}_{\alpha}\rangle^{2} /\langle\hat{Q}_{\alpha}, \hat{Q}_{\alpha}\rangle>0$ for some
% translationally invariant  and 
local observable $\hat{A}$.
\hlik{Here, $\hat{A}$ is traceless and normalized as $0<\lim _{L \rightarrow \infty}\|\hat{A}\|<\infty$ where $\|\hat{A}\|\equiv \sqrt{\braket{\hat{A},\hat{A}}}$  and $L$ denotes the system size.}
%so that it is well-defined in the thermodynamic limit ($L$ denotes the system size).}

\hlik{
If the norm of $\hbO_{\perp}$ does not vanish in the thermodynamic limit, i.e.,
\begin{align}
\lim_{L\rightarrow\infty}\|\hbO_{\perp}\|^{2}>0,
\end{align}
then $\hbO_{\perp}$ is an additional local conserved quantity which should be incorporated into the GGE.
To see this, we first recall that $\hbO_{\perp}$ is a conserved quantity.
Then, the locality of $\hbO_{\perp}$ follows from the identity 
\begin{align}
\frac{\braket{\hat{A},\hbO_{\perp}}^{2}}{\braket{\hbO_{\perp},\hbO_{\perp}}}
=
\braket{\hbO_{\perp},\hbO_{\perp}}=\|\hbO_{\perp}\|^2
>0,
\label{vanish}
\end{align}%}
where we have used $\braket{\hbO_{\perp},\hbO_{\perp}}=\braket{\hat{A},\hbO_{\perp}}$~\cite{PhysRevLett.124.040603}.
}

\hlf{
We remark that the opposite is not true (see Supplemental Material of Ref.~\cite{PhysRevLett.124.040603}). Namely, $\lim_{L\rightarrow\infty}\|\hbO_{\perp}\|=0$ does not necessarily mean the absence of the additional local conserved quantities and  more over, $\lim_{L\rightarrow\infty}\hbO_{\perp}=0$ as an operator.
}
%So we cannot judge whether there are additional local conserved quantities which should be incorporated into the GGE or not. 

%################################################################################################
%################################################################################################
\hlf{
\section{How to detect the necessity of noncommutative conserved quantities}\label{detection}
Now we discuss how to judge whether noncommutative conserved quantities
need to be involved in the GGE or not
when we have only commutative set of conserved quantities $\{\hat{Q}_{\alpha}^{\mathrm{C}}\}$:
$[\hat{Q}^\mathrm{C}_\alpha,\hat{H}]=[\hat{Q}^\mathrm{C}_\alpha,\hat{Q}^\mathrm{C}_\beta]=0$
($\forall\alpha,\beta$).
\hlf{Let $\hat{A}$ be the local observable of interest.}
We remark that $[\hbO,\hat{Q}^\mathrm{C}_\alpha]\neq0$ in general
while $[\hbO,\hat{H}]=0$ follows from the definition.
\hlf{From now on,
\hlik{we suppose that $\hat{Q}_{\alpha}^{\mathrm{C}}$'s are orthonormal and $\hat{H}$ is written as a} linear combination of $\hat{Q}_{\alpha}^{\mathrm{C}}$'s.}
In this situation, we can decompose $\hbO$ into two parts:
\begin{align}
\hbO=\hbO\ \!\!^{\mathrm{C}}+\hbO\ \!\!^{\mathrm{NC}},\label{cncdecomp}
\end{align}%
where $\hbO\ \!\!^{\mathrm{C}}$ and $\hbO\ \!\!^{\mathrm{NC}}$
does and does not commute with $\{\hat{Q}_{\alpha}^{\mathrm{C}}\}$, respectively.
More explicitly, these are defined by
$\hbO\ \!\!^{\mathrm{C}}\equiv \sum_{\{q_{\alpha}\}} \hat{P}_{\{q_{\alpha}\}}\hbO \hat{P}_{\{q_{\alpha}\}}$
and $\hbO\ \!\!^{\mathrm{NC}}\equiv \hbO-\hbO\ \!\!^{\mathrm{C}}$,
where $\hat{P}_{\{q_{\alpha}\}}$ represents the projection operator
onto the simultaneous eigenspace for  $\{\hat{Q}_{\alpha}^{\mathrm{C}}\}$
in which  $\hat{Q}_{\alpha}^{\mathrm{C}}$ have eigenvalues  $q_{\alpha}$, respectively.
Note that $\hbO\ \!\!^{\mathrm{C}}$ and $\hbO\ \!\!^{\mathrm{NC}}$ are orthogonal to each other $\braket{\hbO\ \!\!^{\mathrm{C}},\hbO\ \!\!^{\mathrm{NC}}}$=0.
}

\hlik{
If $\hbO\ \!\!^{\text{NC}}$ does not vanish in the thermodynamic limit,
then the CGGE can fail and the NCGGE is necessary %depending on the initial state.
as follows.
%We consider the CGGE \hlf{in a situation where $\hbO\ \!\!^{\mathrm{NC}}$ exists}.
Let us use the commutative orthogonal set of the conserved quantities $\{\hat{Q}_{\alpha}^{\text{C}}\}$ for the operator projection.
The diagonal component of the perpendicular term becomes 
$\hbO_{\perp}=\hbO\ \!\!^{\text{C}}_{\perp}+\hbO\ \!\!^{\text{NC}}$
where $\hbO\ \!\!^{\text{C}}_{\perp}=\hbO\ \!\!^{\text{C}}-\sum_{\alpha} p_{A \alpha}^{\text{C}} \hat{Q}_{\alpha}^{\text{C}}$
 and 
 $p_{A \alpha}^{\text{C}}=\braket{p_{A \alpha}^{\text{C}},\hat{Q}_{\alpha}^{\text{C}}}/\braket{\hat{Q}_{\alpha}^{\text{C}},\hat{Q}_{\alpha}^{\text{C}}}$
 and the noncommutative part is unchanged.
 }

\hlik{
Even if we do the best in the commutative part so that the commutative set $\{\hat{Q}_{\alpha}^{\text{C}}\}$ is enough and $\lim_{L\rightarrow \infty }\hbO\ \!\!^{\text{C}}_{\perp}=0$,
we still have the CGGE error as
\begin{align}
\lim_{L\rightarrow\infty}\left(\braket{\hat{A}}_{\text{LT}}-\braket{\hat{A}}_{\text{CGGE}}\right)=\lim_{L\rightarrow\infty} \braket{\hbO\ \!\!^{\text{NC}}}_{\text{ini}},\label{cggeerrorofnc}
\end{align}%
where $\braket{\hat{A}}_{\text{CGGE}}$ is the CGGE expectation value of $\hat{A}$ and we use $\braket{\hbO\ \!\!^{\text{NC}}}_{\text{CGGE}}=0$.
Equation~\eqref{cggeerrorofnc}
highlights the failure of the CGGE and
the necessity of the NCGGE depending on the initial state.
}
%From (\ref{cggeerrorofnc}), we can see when  $\braket{\hbO\ \!\!^{\text{NC}}}_{\text{ini}}$ do not vanish in the thermodynamic limit, the CGGE fails and we have to consider the NCGGE.
% This means if $\hbO\ \!\!^{\text{NC}}$ do not vanish in the thermodynamic limit, the CGGE fails depending on the initial state.

%######
%If $\hbO$ consists of a noncommutative set of conserved quantities even in the thermodynamic limit, the GGE needs to involve them and become an NCGGE.

%\hlf{ Recently, the explicit forms of all the local conserved quantities of XYZ chain was found. We can immediately apply the above argument to the XYZ chain.}
%As we remarked before, the NCGGE enlarges the possible set of conserved quantities and gives a more accurate description of the long-time average in general.
%Since the role of noncommutativity is not apparent in this general discussion,
%we will discuss a concrete model below.

\section{Calculation of generalized temperatures for one-body NCGGE}\label{generalized temperature}
We study the explicit form of the generalized temperatures $\lambda_k$ and $ \omega_k$.
The density matrix of  the one-body NCGGE is
\begin{align}
  \hat{\rho}_{\text{1NC}}
  &=
  \frac{1}{Z_{\text{1NC}}}
  \exp\left[- \sum_{k=-\pi}^{\pi}(\lambda_k\I_k+\omega_k\J_k)\right],
\end{align}
where
$
  Z_{\text{1NC}}
  =
  \tr
  e^{- \sum_{k}(\lambda_k\I_k+\omega_k\J_k)}
$.
To make the density matrix Hermitian, we impose $\omega_{k}^*=\omega_{-k}$ because of $\J_{k}^{\dag}=\J_{-k}$. 
We note that $\lambda_{k}$ is real since $\I_{k}^{\dag}=\I_{k}$.

%一般化温度が厳密に求まることを述べる
The generalized temperatures $\lambda_{k}$ and $\omega_k$ are uniquely and explicitly determined from the conditions
$\braket{\I_{k}}_{\text{ini}}=\text{Tr}[\hat{\rho}_{\text{1NC}}\I_{k}]$
and $\braket{\J_{k}}_{\text{ini}}=\text{Tr}[\hat{\rho}_{\text{1NC}}\J_{k}]$
We note that
$\hat{\rho}_{\text{1NC}}$ consists of product of the following $(k,-k)$-subspace operators
\begin{align}
  \hat{X}_k&\equiv
  \lambda_k\I_k+\omega_k\J_k+\lambda_{-k}\I_{-k}+\omega_{-k}\J_{-k}
  \nonumber\\
  &=
  \begin{pmatrix}
    \hat{c}_k^\dag &\hat{c}_{-k}^\dag
  \end{pmatrix}
  \begin{pmatrix}
    \lambda_k & \omega^*_k\\
    \omega_k & \lambda_{-k}
  \end{pmatrix}
  \begin{pmatrix}
    \hat{c}_k \\
    \hat{c}_{-k}
  \end{pmatrix}.
 \label{xk}
\end{align}
Then the density matrix of the one-body NCGGE can be written as 
$
\hat{\rho}_{\text{1NC}}
=
Z_{\text{1NC}}^{-1} \prod_k e^{-\hat{X}_k}
$.
We diagonalize the matrix in Eq.~\eqref{xk}.
The hermitian matrix can be written by the liner combination of Pauli matrices and Identity matrix
\begin{align}
  \begin{pmatrix}
    \lambda_{k} & \omega^*_{k}\\
    \omega_{k} & \lambda_{-k}
  \end{pmatrix}
  &=
  \bar{\lambda}_{k}I+\mathrm{Re}\omega_{k} \sigma_x+\mathrm{Im}\omega_{k} \sigma_y+\Delta \lambda_{k}\sigma_z
  \\
  &=
  \bar{\lambda}_{k}I+\bm{a}_{k}\cdot\bm{\sigma}
  =
  \bar{\lambda}_{k}I+a_{k}\bm{n}_{k}\cdot\bm{\sigma},
\end{align}
where 
$\bar{\lambda}_{k}=\frac{1}{2}(\lambda_{k}+\lambda_{-k})$
,
$\Delta\lambda_{k}=\frac{1}{2}(\lambda_{k}-\lambda_{-k})$
,
$\bm{a}_{k}=
(
\mathrm{Re}\omega_{k}, \mathrm{Im}\omega_{k},   \Delta \lambda_{k}
)$,
and
$
\bm{\sigma}
=
(
\sigma_x,\sigma_y,\sigma_z
)$.
 We define the unit vector
$\bm{n}_{k}=\bm{a}_{k}/a_k$,
where $a_{k}=|\bm{a}_{k}|$.
\\

We rotate $\bm{a}_{k}\cdot\bm{\sigma}$
to
$\sigma_z$
by a unitary transformation
\begin{align}
U_{k}^{\dag}\bm{n}_{k}\cdot\bm{\sigma}U_{k}=\sigma_z.
\end{align}%}
Thus we obtain 
\begin{align}
U_{k}^{\dag}
 \begin{pmatrix}
    \lambda_{k} & \omega^*_{k}\\
    \omega_{k} & \lambda_{-k}
  \end{pmatrix}
U_{k}
=
 \begin{pmatrix}
    \bar{\lambda}_{k}+a_{k} &0\\
   0&\bar{\lambda}_{k}-a_{k} 
  \end{pmatrix}.
\end{align}%}
An explicit form of the unitary transformation is given by
\begin{align}
  U_{k}^{\dag}
  &=
  e^{i\frac{\phi_{k}}{2}\sigma_{z}}
e^{i\frac{\theta_{k}}{2}\sigma_{y}} \\
  &=
  \begin{pmatrix}
    e^{\frac{i}{2}\phi}\cos(\theta/2) & e^{-\frac{i}{2}\phi}\sin(\theta/2)\\
    -e^{\frac{i}{2}\phi}\sin(\theta/2) & e^{-\frac{i}{2}\phi}\cos(\theta/2)
  \end{pmatrix},
\end{align}
where $\theta_{k}$ and $\phi_{k}$ are the polar and azimuthal angles of $\bm{n}_{k}$.
%, which are determined from the initial conditions.
%
The corresponding transformation of the annihilation operators is
\begin{align}
  \begin{pmatrix}
    d_k\\
    d_{-k}
  \end{pmatrix}
  &=
  U_{k}^{\dag}
  \begin{pmatrix}
    c_{k}\\
    c_{-k}
  \end{pmatrix}.
\end{align}
%
\begin{comment}
&=
  \begin{pmatrix}
    e^{\frac{i}{2}\phi_{k}}\cos(\theta_{k}/2) c_{k} +e^{-\frac{i}{2}\phi_{k}}\sin(\theta_{k}/2) c_{-k}\\
    -e^{\frac{i}{2}\phi_{k}}\sin(\theta_{k}/2) c_{k} + e^{-\frac{i}{2}\phi_{k}}\cos(\theta_{k}/2) c_{-k}
  \end{pmatrix}
\end{comment}
%
Note that the unitary transformation preserves the anti-commutation relations
\begin{align}
  \{d_{\sigma k},d_{\rho k}^\dag \}
  =
  (U_{k}^\dag U_{k})_{\sigma\rho}
  =
  \delta_{\sigma\rho},
\end{align}
where $\sigma$ and $\rho=\pm 1$.
Then, $\hat{X}_{k}$ becomes
\begin{align}
  \hat{X}_{k}
 & =
  (\bar{\lambda}_{k}+a_{k})d_k^\dag d_{k}
  +
  (\bar{\lambda}_{k}-a_{k})d_{-k}^\dag d_{-k}
  \nonumber\\
  &=
 \eta_{k}\Id_{k}
  +
  \eta_{-k}\Id_{-k},
\end{align}
where $\Id_{k}=d^{\dag}_{k}d_{k}$ is the rotated conserved quantitiy and $\eta_{\pm k}= \bar{\lambda}_{k} \pm a_{k}$.
The density matrix is then diagonalized in the $d_{k}$-basis as
\begin{align}
\hat{\rho}_{\text{1NG}}
=
\frac{1}{Z_{\text{1NG}}}
 \prod_{k=-\pi}^{\pi}
  \exp\left(-\eta_k \Id_{k}\right).
\end{align}%
Note that $\Id_{k}$ commutes with each other $[\Id_{k},\Id_{q}]=0$,
and $\lambda_{k}$ and $\omega_k$ are written as 
\begin{align}
\lambda_{\pm k}=\bar{\eta}_{k}\pm  \Delta\eta_{k}\cos\theta_{k},
\label{lambda}
\\
 \omega_{\pm k}= \Delta\eta_{k}e^{\pm i\phi_{k}}\sin\theta_{k}
 \label{omega},
\end{align}
where $\bar{\eta}_{k}=(\eta_{k}+\eta_{-k})/2, \Delta\eta_{k}=(\eta_{k}-\eta_{-k})/2$.
The determining equations for $\theta_{k}, \phi_{k}$, and  $\eta_{\pm k}$ are
\begin{align}
\braket{\I_{\pm k}}_{\text{ini}}
&=
\frac{\cos^{2}\theta_{k}/2}{1+e^{\eta_{k}}}+\frac{\sin^{2}\theta_{k}/2}{1+e^{\eta_{-k}}},
\\
\braket{\J_{\pm k}}_{\text{ini}}
&=
\frac{e^{\mp i\phi_{k}}}{2}
\sin\theta_{k}
(
\frac{1}{1+e^{\eta_{k}}}-\frac{1}{1+e^{\eta_{-k}}}
).
\end{align}
Solving these equations, we have 
\begin{align}
e^{i\phi_{k}}
&=
-|\braket{\J_{k}}_{\text{ini}}|/\braket{\J_{k}}_{\text{ini}}\label{expphi}
,\\
\cos \theta_{k}
&=
-
\frac{\braket{\I_{k}}_{\text{ini}}-\braket{\I_{-k}}_{\text{ini}}}{\sqrt{(\braket{\I_{k}}_{\text{ini}}-\braket{\I_{-k}}_{\text{ini}})^{2}+4|\braket{\J_{k}}_{\text{ini}}|^{2}}}\label{costheta}
,\\
\sin \theta_{k}
&=
\frac{2|\braket{\J_{k}}_{\text{ini}}|}{\sqrt{(\braket{\I_{k}}_{\text{ini}}-\braket{\I_{-k}}_{\text{ini}})^{2}+4|\braket{\J_{k}}_{\text{ini}}|^{2}}}\label{sintheta}
,\\
\braket{{\Id_{\pm k}}}_{\text{ini}}&=\frac{1}{1+e^{\eta_{\pm k}}}
\nonumber\\
&= \frac{\braket{\I_{k}}_{\text{ini}}+\braket{\I_{-k}}_{\text{ini}}}{2}
\nonumber\\
\mp&
\sqrt{(\braket{\I_{k}}_{\text{ini}}-\braket{\I_{-k}}_{\text{ini}})^{2}/4+|\braket{\J_{k}}_{\text{ini}}|^{2}}.
\label{idexp}
\end{align}

Using (\ref{expphi}- \ref{idexp}), the explicit forms of $\phi_{k}, \theta_{k}$, and $\eta_{k}$ are obtained as 
\begin{align}
\phi_{k}
&=
\pi-\text{arg}\braket{\J_{k}}_{\text{ini}},
\label{phi}
\\
\theta_{k}
&=
-\tan^{-1}[2|\!\braket{\J_{k}}_{\text{ini}}\!|/(\braket{\I_{k}}_{\text{ini}}-\braket{\I_{-k}}_{\text{ini}})],
\label{theta}
\\
\eta_{k}
&=
\log(\frac{1}{\braket{{\Id_{k}}}_{\text{ini}}}-1).
\end{align}
Then, we obtain the generalized temperatures $\lambda_k$ and $ \omega_k$ from Eqs.~\eqref{lambda} and \eqref{omega}.

\section{determination of generalized temperature for trigonal NCGGE}\label{tnc temp}
We discuss the generalized temperatures of the trigonal NCGGE.
For this purpose in this section, we introduce an abuse of notation $\Id_K$
for $\Id_k$,
where $K$ is an integer satisfying 
\begin{align}
k=2\pi K/L \qquad \mod 2\pi.
\end{align}
%密度行列を提示
Then the density matrix of the  trigonal NCGGE is 
\begin{align}
 \hat{\rho}_{\text{tNC}}
  &=
  \frac{1}{Z_{\text{tNC}}}
  \exp\left(- \sum_{K}\tilde{\eta}_K\Id_K-\sum_{K}\Lambda_{K}\Id_K\Id_{K+1}\right)
  \nonumber\\
  &=
  \frac{1}{Z_{\text{tNC}}}
   \prod_{K=0}^{L-1}
  T_{K}(\Id_{K},\Id_{K+1}),
\end{align}%
where
   $ T_{K}(\Id_{K},\Id_{K+1})$ is the transfer matrix operator
   \begin{align}
   &T_{K}(\Id_{K},\Id_{K+1})
   \nonumber\\
  & =
   \exp
   [
   -(
   \Lambda_{K}\Id_{K}\Id_{K+1}
   +(\tilde{\eta}_K\Id_{K}+\tilde{\eta}_{K+1}\Id_{K+1})
   )/2
   ].
   \end{align}%}
% We can  calculate the generalized temperatures  $\tilde{\eta}_K, \Lambda_{K}$ iteratively by  the transfer matrix method like one-dimensional Ising model.
 %転送行列の方法を大雑把に説明
 In analogy with the Ising model in one dimension,
 we define the transfer matrix as
 \begin{align}
 T_{K}
 &=
  \begin{pmatrix}
 T_{K}(1,1) & T_{K}(1,0)\\
 T_{K}(0,1) & T_{K}(0,0)
 \end{pmatrix}
 \nonumber
 \\
 =&
 \begin{pmatrix}
 e^{-\Lambda_{K}-(\eta_{K}+\eta_{K+1})/2} & e^{-\eta_{K}/2}\\
 e^{-\eta_{K+1}/2} & 1
 \end{pmatrix}.
 \end{align}
 By using the transfer matrix,
we can calculate
the partition function and the expectation values of each conserved quantity
in the trigonal NCGGE as
\begin{align}
 Z_\text{tNC}
 &=
 (T_{0}T_{1}\ldots T_{L-1})_{00}
 +
 (T_{0}T_{1}\ldots T_{L-1})_{11},
 \\
 \braket{\Id_{K}}_{\text{tNC}}
 &=
 \frac{(T_{K}T_{K+1}\ldots T_{L-1}T_{0}T_{1}\ldots T_{K-1})_{11}}{Z_{\text{tNC}}},
 \\
 \braket{\Id_{K}\Id_{K+1}}_{\text{tNC}}
& =
 \frac{(T_{K})_{11}
 (T_{K+1}\ldots T_{L-1}T_{0}T_{1}\ldots T_{K-1})_{11}}{Z_{\text{tNC}}}.
  \end{align}%
We remark that the right-hand sides of these equations 
can be numerically evaluated in polynomial times rather than exponential ones.

The determining equations for the generalized temperatures
are $\braket{\Id_{K}}_{\text{tNC}}=\braket{\Id_{K}}_{\text{ini}}$ and $\braket{\Id_{K}\Id_{K+1}}_{\text{tNC}}=\braket{\Id_{K}\Id_{K+1}}_{\text{ini}}$,
which are equivalent to the following self-consistent equations
for $T_K$:
\begin{align}
 (T_{K})_{11}=
\frac{Z_{\text{tNC}}\braket{\Id_{K}\Id_{K+1}}_{\text{ini}}}{(T_{K+1}\ldots T_{L-1}T_{0}\ldots T_{K-1})_{11}},
\label{t00}\\
(T_{K})_{10}=
\frac{Z_{tNC}(\braket{\Id_{K}}_{\text{ini}}-\braket{\Id_{K}\Id_{K+1}}_{\text{ini}})}{(T_{K+1}\ldots T_{L-1}T_{0}\ldots T_{K-1})_{01}}.
\label{t01}
\end{align}
By iteratively calculating $T_K$,
we obtain the generalized temperatures $\eta_{k}$ and $\Lambda_{k}$.

%##############################

 \begin{figure}[tb]
 \centering
 \includegraphics[width=\columnwidth]{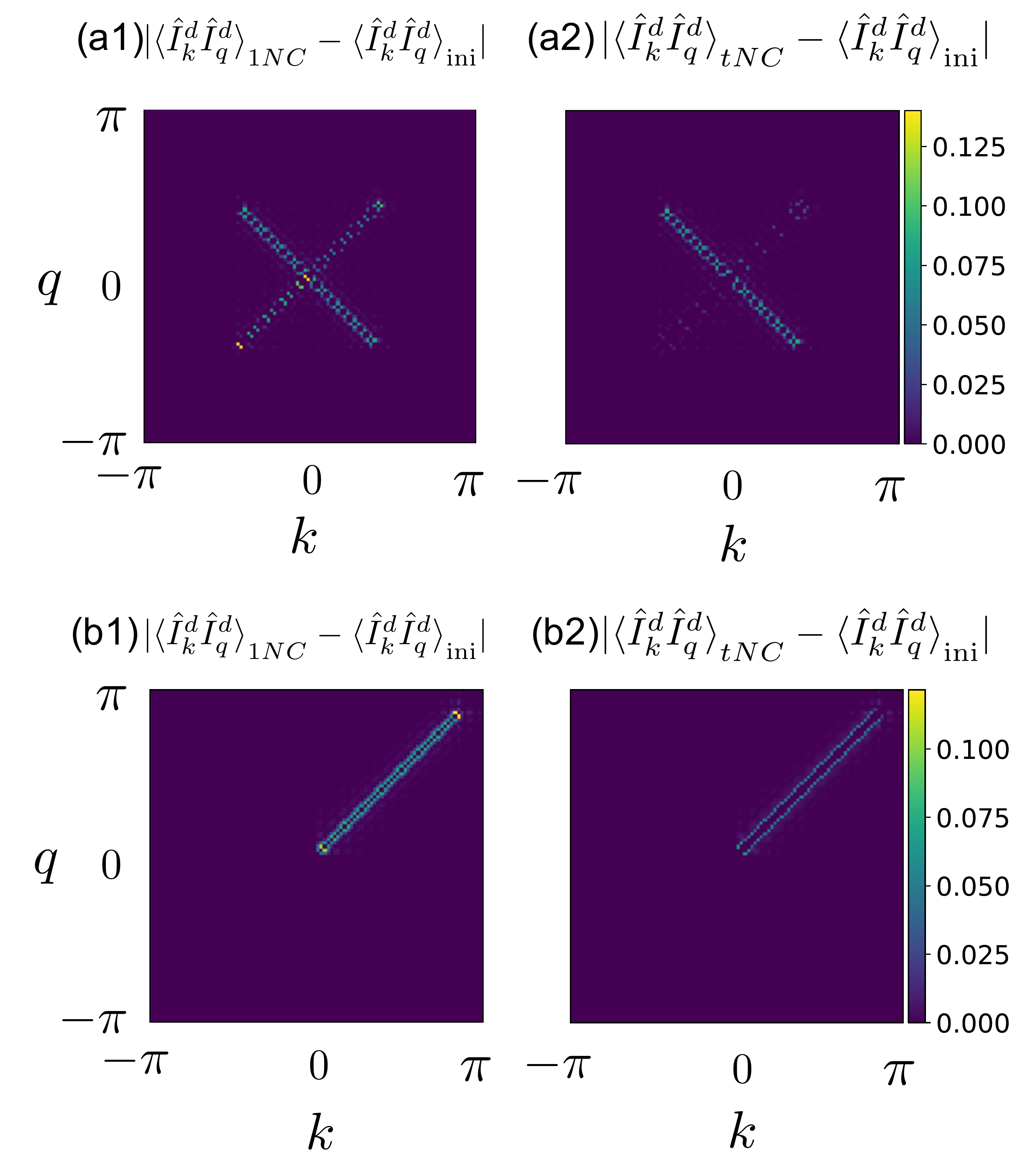}
 \caption{
The differences of the expectation value of $\Id_{k}\Id_{q}$ from the initial state expectation value are plotted.
The one-body NCGGE case $|\braket{\Id_{k}\Id_{q}}_{\text{ini}}-\braket{\Id_{k}\Id_{q}}_{1NC}|$ with the initial state $\ket{\psi_{\text{ini}}^{A}}$ is (a1) and that  with the initial state $\ket{\psi_{\text{ini}}^{B}}$ is  (b1).
The trigonal  NCGGE case $|\braket{\Id_{k}\Id_{q}}_{\text{ini}}-\braket{\Id_{k}\Id_{q}}_{tNC}|$ with the initial state $\ket{\psi_{\text{ini}}^{A}}$ is (a2) and that  with the initial state $\ket{\psi_{\text{ini}}^{B}}$ is  (b2).
The color bars are common in upper panels and in lower panels respectivelly.
The system size is $L=100$ and the particle number is $N=30$.
The initial hard wall box is the size of   $L_{\text{ini}}=70$. 
} 
\label{ndnd-vs}
\end{figure}

 \begin{figure}[tb]
 \centering
 \includegraphics[width=\columnwidth]{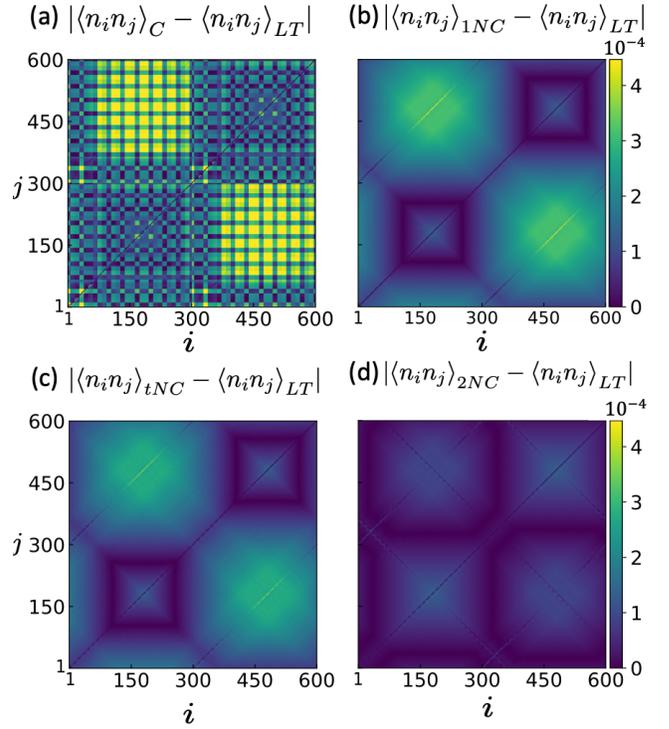}
\caption{
Error of GGEs $|\braket{\hat{n}_{i}\hat{n}_{j}}_{\mathrm{GGE}}-\braket{\hat{n}_{i}\hat{n}_{j}}_{\mathrm{LT}}|$
for the density-density correlation between sites $i$ and $j$
calculated with the (a) CGGE, (b) one-body NCGGE, (c) trigonal NCGGE, (d) two-body NCGGE.
In all panels, $L=600$, $L_{\text{ini}}=360$, and $N=120$.
and we use the initial state $\ket{\psi_{\text{ini}}^{\text{A}}}$,
and implicitly assume the normal ordering for $\hat{n}_{i}\hat{n}_{j}$(see footnote [57]).
} 
\label{errorground}
\end{figure}

\begin{figure}[tb]
 \centering
 \includegraphics[width=\columnwidth]{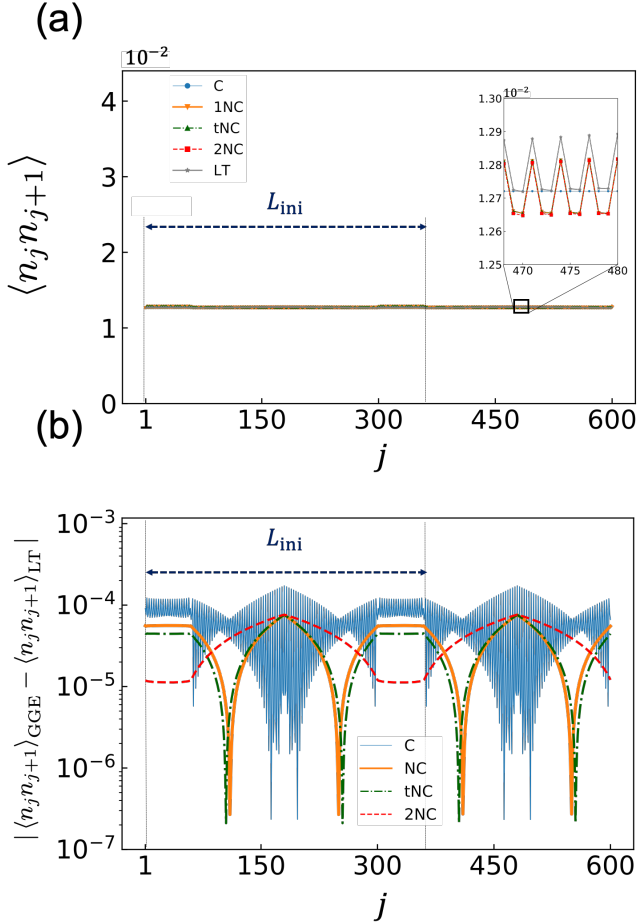}
\caption{
\hl{
(a) Expectation values of local density-density correlation of  $\braket{n_jn_{j+1}}$ in the CGGE, one-body NCGGE, trigonal NCGGE, two-body NCGGE, and long-time average with the initial state $\ket{\psi_{\text{ini}}^{\text{A}}}$ and $L=600$, $L_{\text{ini}}=360$, and $N=120$. There are no characteristic peaks which cannot be explained by the CGGE  such as in the case of $\ket{\psi_{\text{ini}}^{\text{B}}}$ .
(b) Error of GGEs $|\braket{\hat{n}_{j}\hat{n}_{j+1}}_{\mathrm{GGE}}-\braket{\hat{n}_{j}\hat{n}_{j+1}}_{\mathrm{LT}}|$ for the local density-density correlation between sites $j$ and $j+1$.
}
} 
\label{localground}
\end{figure}

\section{$(k,-k)$-subspace NCGGE}
Though we cannot easily take two-body operator into the GGE, $\I_{k}\I_{-k}$  can be easily taken into the GGE because we can diagonalize the density matrix in each (k,-k) subspace as one-body NCGGE.
However, when we use the initial state of the product of the single particle state, the result is the same as one-body NCGGE.

Note that  $\I_{k}\I_{-k}$ is invariant in the unitary transformation, or $\I_{k}\I_{-k}=\Id_{k}\Id_{-k}$.
We call the GGE with the conserved quantities $\I_{k}$, $\J_{k}$, $\I_{k}\I_{-k}$ as the (k,-k) subspace GGE(sGGE).
The density matrix of the sGGE is 
$
 \hat{\rho}_{\text{sNG}}
 =
\frac{1}{Z_{\text{sNG}}}e^{- \sum_{0<k<\pi}\hat{Y}_k}
 $,
 where $Z_{\text{sNG}}=\tr e^{- \sum_{0<k<\pi}\hat{Y}_k}$ is the partition function and
\begin{align}
 \hat{Y}_k
   &=
 \lambda_k\I_k+\omega_k\J_k+\lambda_{-k}\I_{-k}+\omega_{-k}\J_{-k}
 +
 \Gamma_k\I_k\I_{-k}.
 \end{align}%}
 We rotate the basis as in the one-body NCGGE. The rotated form of $ \hat{Y}_k$ by $U_{k}$ is
 
 \begin{align}
  \hat{Y}_k
   &=
   \eta_{k}\Id_{k}
  +
  \eta_{-k}\Id_{-k}
   +
 \Gamma_k\Id_k\Id_{-k}.
 \end{align}%}
 The definitions of these symbols are the same as the one-body NCGGE case.
 The initial state expectation value of the conserved quantities are
 \begin{align}
 \braket{\I_{\pm k}}_{\text{ini}}
 &=
 \cos^{2}\frac{\theta_{k}}{2}x_{\pm k}+\sin^{2}\frac{\theta_{k}}{2}x_{\mp k}+y_{k},
 \\
  \braket{\J_{\pm k}}_{\text{ini}}
 &=
 \frac{e^{\mp i\phi_{k}}}{2}
\sin\theta_{k}
(
x_{k}-x_{-k}
),
\\
\braket{\I_{k}\I_{- k}}_{\text{ini}}
&=
y_{k},
 \end{align}
 where 
 $
 x_{ k}=e^{-\eta_{k}}/z_{k}
 ,\  
 y_{k}=e^{-(\eta_{k}+\eta_{-k}+\Gamma_{k})}/z_{k}
,\ 
 z_{k}
 =
1+e^{-\eta_{k}}+e^{-\eta_{-k}}+e^{-(\eta_{k}+\eta_{-k}+\Gamma_{k})}
 $.
 Solving these equations for $x_{\pm k}, z_{k}$, we have 
 \begin{align}
 x_{\pm k}
 &=
 \frac{ \braket{\I_{ k}}_{\text{ini}}+\braket{\I_{-k}}_{\text{ini}}}{2}-\braket{\I_{k}\I_{- k}}_{\text{ini}}
 \nonumber\\
 &
 \pm
 \sqrt{(\braket{\I_{k}}_{\text{ini}}-\braket{\I_{-k}}_{\text{ini}})^{2}/4+|\braket{\J_{k}}_{\text{ini}}|^{2}},
 \\
 z_{k}
& =
 \frac{1}{1+\braket{\I_{k}\I_{- k}}_{\text{ini}}-\braket{\I_{k}}_{\text{ini}}-\braket{\I_{k}}_{\text{ini}}}.
 \end{align}%}
 From this, we can see the rotation angle $ \phi_{k}, \theta_{k}$ is the same as the one-body NCGGE\eqref{phi},\eqref{theta}.
 $\braket{{\Id_{\pm k}}}_{\text{ini}}$ is also the same as \eqref{idexp}.
 We can calculate $\eta_{k}$ as $\eta_{k}=-\log({x_{k}z_{k}})$.
 Therefore we can calculate the generalized temperature $\lambda_{k}, \omega_{k}$ with	 \eqref{lambda}, \eqref{omega}, \eqref{phi},\eqref{theta}.
 \hlf{$\Gamma_{k}$} is obtained as 
 \begin{align}
 \Gamma_{k}
 =
 \log{\frac{x_{k}x_{-k}z_{k}}{y_{k}}}.
 \end{align}

We can calculate the explicit formula of the generalized temperatures of (k,-k) subspace NCGGE because $\Id_{k}\Id_{-k}$ is the operator which acts on the k,-k subspace.
The value of $\theta_{k}$ and $\phi_{k}$ are not affected wether $\Id_{k}\Id_{-k}$ is used in the GGE or not.
Note that  $\Id_{k}\Id_{-k}$ is invariant in the unitary transformation, i.e.  $\Id_{k}\Id_{-k}=\I_{k}\I_{-k}$.

When the initial state is the product of the single particle state, there is no improvement in (k,-k) subspace NCGGE from the one-body NCGGE.
This is because 
$\braket{\Id_{k}\Id_{-k}}_{\text{1NC}}=\braket{\I_{k}}_{\text{ini}}\braket{\I_{-k}}_{\text{ini}}-|\braket{\J_{k}}_{\text{ini}}|^{2}$ and $\braket{\Id_{k}\Id_{-k}}_{\text{sNC}}=\braket{\I_{k}\I_{-k}}_{\text{ini}}$ and we can easily show
\begin{align}
\braket{\I_{k}\I_{-k}}_{\text{ini}}=\braket{\I_{k}}_{\text{ini}}\braket{\I_{-k}}_{\text{ini}}-|\braket{\J_{k}}_{\text{ini}}|^{2},
\end{align}%}
 when the initial state is the product of the single particle state.
 From this, we can see the expectation value of the conserved quantities $\Id_{k}\Id_{-k}$ is the same in the one-body NCGGE and  (k,-k) subspace NCGGE  when the initial state is the product of the single particle state.
The difference of the fitting of the conserved quantities in the two NCGGE is only the fitting of the $\Id_{k}\Id_{-k}$.
Therefore the expectation value of any observables in the one-body NCGGE and the (k,-k) subspace NCGGE is the same when the initial state is the product of the single particle state.

The initial state used in this paper  is the  product of the single particle state. Thus we do not use the  (k,-k) subspace NCGGE because the result is the same  in the one-body NCGGE.

\section{fitting of two-body conserved quantities in one-body and trigonal NCGGEs}\label{2fit}
We study how much the two-body conserved quantities $\Id_{k}\Id_{q}$ are fit by the one-body NCGGE or trigonal NCGGE.
We plot $|\braket{\Id_{k}\Id_{q}}_{\text{ini}}-\braket{\Id_{k}\Id_{q}}_\text{1NC}|$ with the initial state $\ket{\psi_{\text{ini}}^{A}}$ in Fig.~\ref{ndnd-vs}(a1) and with the initial state $\ket{\psi_{\text{ini}}^{B}}$ in Fig.~\ref{ndnd-vs}(b1).
We also plot $|\braket{\Id_{k}\Id_{q}}_{\text{ini}}-\braket{\Id_{k}\Id_{q}}_\text{tNC}|$ with the initial state $\ket{\psi_{\text{ini}}^{A}}$ in Fig.~\ref{ndnd-vs}(a2) and with the initial state $\ket{\psi_{\text{ini}}^{B}}$ in Fig.~\ref{ndnd-vs}(b2).
In both initial state case, We find that most deviations reside around
the diagonal ($k=q$) and anti-diagonal ($k=-q$) components in one-body NCGGE case.
The trigonal NCGGE is also made from the two-body conserved quantities $\Id_k\Id_{k+\Delta k}$.
Comparing  Fig.~\ref{ndnd-vs}(a1,b1) and Fig.~\ref{ndnd-vs}(a2,b2), we can see the fitting of the trigonal components of the two-body conserved quantities are improved from the one-body NCGGE.

%###################33
\section{density-density correlation for ground initial state $\ket{\psi_\text{ini}^\text{A}}$} \label{ground initial}
 We show the expectation values of the density-density correlation in the CGGE and the NCGGEs for the ground initial state $\ket{\psi_{\text{ini}}^\text{A}}$ \hlf{in Fig.~\ref{errorground}}.
  We evaluate the error $|\braket{\hat{n}_{i}\hat{n}_{j}}_\text{GGE}-\braket{\hat{n}_{i}\hat{n}_{j}}_{\mathrm{LT}}|$, where GGE means the C-, one-body NC-, trigonal NC-, and two-body NC-GGE.
  We find that the more conserved quantities are used, the more accurate the GGE becomes.
  For a quantitative comparison of local observables,
we plot the expectation values of $\hat{n}_{j}\hat{n}_{j+1}$ in Fig.~\ref{localground}(a).
There are no characteristic peaks that survives in the thermodynamic limit, and the CGGE is thus accurate
for density-density correlation with any pair of sites
unlike the excited initial state $\ket{\psi_{\text{ini}}^\text{B}}$.
The difference between the two-body NCGGE and the long-time average is due to the accidental degeneracy.

\end{document}